\documentclass[twocolumn, 11pt]{aastex63}
\usepackage{epsfig,xcolor}
\usepackage{natbib}
\usepackage[normalem]{ulem}
\usepackage{comment}
\usepackage{appendix}
\usepackage{mwe}

\begin{document}

\title{A Machine Learning Approach to Enhancing \textit{eROSITA} Observations}

\author{John Soltis}
\affiliation{Department of Physics \& Astronomy, Johns Hopkins University, Baltimore, MD 21218, USA}
\author{Michelle Ntampaka}
\affiliation{Space Telescope Science Institute, Baltimore, MD 21218, USA}
\affiliation{Department of Physics \& Astronomy, Johns Hopkins University, Baltimore, MD 21218, USA}
\author{John F. Wu}
\affiliation{Space Telescope Science Institute, Baltimore, MD 21218, USA}
\affiliation{Department of Physics \& Astronomy, Johns Hopkins University, Baltimore, MD 21218, USA}

\author{John ZuHone}
\affiliation{Center for Astrophysics $\mid$ Harvard \& Smithsonian, 60 Garden Street, Cambridge, MA 02138, USA}

\author{August Evrard}
\affiliation{Departments of Physics and Astronomy and Leinweber Center for Theoretical Physics, University of Michigan, Ann Arbor, MI 48109, USA}
\author{Arya Farahi}
\affiliation{Departments of Statistics and Data Science, University of Texas at Austin, Austin, TX 78705, USA}
\author{Matthew Ho}
\affiliation{McWilliams Center for Cosmology, Department of Physics, Carnegie Mellon University, Pittsburgh, PA 15213, USA}
\affiliation{NSF AI Planning Institute for Physics of the Future, Carnegie Mellon University, Pittsburgh, PA 15213, USA}
\author{Daisuke Nagai}
\affiliation{Department of Physics, Yale University, New Haven, CT 06520, USA}

\begin{abstract}
    The \textit{eROSITA} X-ray telescope, launched in 2019, is predicted to observe roughly 100,000 galaxy clusters. Follow-up observations of these clusters from \textit{Chandra}, for example, will be needed to resolve outstanding questions about galaxy cluster physics. Deep \textit{Chandra} cluster observations are expensive and follow-up of every \textit{eROSITA} cluster is infeasible, therefore, objects chosen for follow-up must be chosen with care. To address this, we have developed an algorithm for predicting longer duration, background-free observations based on mock \textit{eROSITA} observations. We make use of the hydrodynamic cosmological simulation \texttt{Magneticum}, have simulated \textit{eROSITA} instrument conditions using \texttt{SIXTE}, and have applied a novel convolutional neural network to output a deep \textit{Chandra}-like ``super observation'' of each cluster in our simulation sample. Any follow-up merit assessment tool should be designed with a specific use case in mind; our model produces observations that accurately and precisely reproduce the cluster morphology, which is a critical ingredient for determining cluster dynamical state and core type. Our model will advance our understanding of galaxy clusters by improving follow-up selection and demonstrates that image-to-image deep learning algorithms are a viable method for simulating realistic follow-up observations.\\
\end{abstract}

\section{Introduction}\label{introduction}

Galaxy clusters are the most massive gravitationally bound objects in the Universe. They consist of scores to hundreds to thousands of galaxies in a common dark matter halo. Galaxies and the intra-cluster medium (ICM) form the ordinary baryonic matter component of these structures and emit light across the electromagnetic spectrum, allowing us to observe them. Through brehmsstrahlung, collisional excitation, recombination radiation, and 2-photon emission processes, the ICM produces X-ray photons, allowing for X-ray observations of clusters. Galaxy clusters are an important probe of dark matter \citep[e.g.,][]{Clowe_2006} and are of special interest to cosmologists because they are the high density peaks of the present-day Universe and are sensitive to the underlying cosmological model \citep[see][for a recent review]{Pratt_2019}.

Galaxy clusters' sensitivity to cosmological parameters makes them excellent probes of cosmology. The abundance as a function of mass and redshift provides us with information about the underlying cosmological model, in particular the matter density, $\Omega_m$, and the amplitude of matter fluctuations, $\sigma_8$ \citep{Allen_2011, Kravtsov_Borgani_2012, Pratt_2019}. Mass estimations of a population of clusters with a well-understood selection function can be used to construct a halo mass function \citep[e.g.,][]{Tinker_2008, Bocquet_2016}, which can be used to constrain cosmological models. X-ray observations are especially useful for mass estimation because they provide low scatter proxies of cluster mass \citep[e.g.,][]{Kravtsov_2006}. However, mass estimations of clusters are reliant on mass proxies, which may result in biased estimates of the true mass \citep[e.g.,][]{Nagai_2007, Lau_2013, Nelson_2014, Shi_2016, Biffi_2016, Barnes_2021}. 

The dynamical state of a cluster, which is a function of its mass accretion history, can substantially bias mass estimation \citep[e.g.,][]{Lau_2009, Nelson_2014, Shi_2015}. To accurately correct the level of bias introduced into mass estimates, some understanding of the dynamical state of the clusters is needed. Conveniently, substantial mass accretion history often leaves noticeable signals. It has a measurable impact on the radial density profile of cluster outskirts \citep{Diemer_2014}, the clumpiness of clusters \citep{Diasuke_Lau_2011}, and their morphology \citep{Evrard_1993}. If one can control for the mass accretion rate, mass bias can be reduced.  High angular resolution, long duration, X-ray observations of galaxy clusters can provide information to characterize cluster dynamical state.

Galaxy cluster core astrophysics is also an area of active inquiry. Populations of clusters can be categorized according to the apparent cooling properties of their cores, ranging from cool core clusters to non-cool core clusters \citep{Jones_Forman_1984}. The origins of these different cluster types are not fully understood, with proposed theories requiring revision upon improved observations \citep[see][ and references therein for reviews]{Fabian_1994, McNamara_Nulsen_2012, Inoue_2022}. Detailed X-ray imaging of galaxy cluster cores is necessary to better understand core dynamics.

\textit{eROSITA} (extended ROentgen Survey with an Imaging Telescope Array) \citep{eROSITA_Sciencebook_2012} will provide an all-sky X-ray survey, and is expected to detect $\sim$100,000 clusters \citep{Pillepich_2018}. \textit{eROSITA}'s observations are complimentary to existing observatories, like \textit{Chandra}. Whereas \textit{Chandra}'s high angular resolution observations offer detailed spatial information on individual clusters, \textit{eROSITA}'s all-sky survey allows for a well-modeled selection function of the underlying galaxy cluster population. Improving our understanding of galaxy clusters requires leveraging both instruments, using \textit{eROSITA}'s observed cluster population to discover cluster populations of interest that are suitable for detailed follow-up observations. However, \textit{eROSITA} will provide us with a plethora of potential follow-up candidates, far surpassing the $\sim$10$^{3}$ large extended sources observed by \textit{Chandra} \citep[the Chandra Source Catalog release 2;][]{Chandra_dr2_A,Chandra_dr2_B}. Given \textit{Chandra}'s operating constraints, this increase in the number of observable clusters is too large for us to conduct follow-up observations on any more than a small fraction. Future observers will need to carefully select follow-up candidates from the \textit{eROSITA} survey. 

The enormous disparity in observable galaxy clusters between \textit{Chandra} and \textit{eROSITA} therefore necessitates a follow-up merit assessment tool. In this work, we present a tool that, given an \textit{eROSITA} observation, provides a prediction of a background-free, long-duration, follow-up observation. This tool illustrates deep learning algorithms' suitability for follow-up merit assessment by predicting a high quality observation from a lower quality observation more accurately and precisely than the original observation or a simple non-deep learning prediction method. 

There is a long history of astronomers developing tools to improve the resolution and the signal to noise ratio of their images \citep[e.g.,][]{Richardson_1972,Lucy_1974,Cornwell_1985} and machine learning offers modern techniques for addressing this problem. For any resolution- or signal-boosting tool to be useful, it must appropriately capture relevant galaxy cluster properties, and the relevant cluster properties are dependent on the science use case.  Cluster shape is one example: morphology is an observable indicator of a cluster's mass distribution, ellipticity, and substructure, and is correlated to cluster dynamical state \citep{ Melott_2001, Rasia_2013, Parekh_2015, Lovisari_2017, Chen_2019, Lau_2021} and core type \citep{Santos_2008}.  Cluster shape measurements are also useful for estimating cluster mass \citep{Green_2019}, and for these reasons, morphological parameters such as cluster surface brightness concentration \citep{Santos_2008}, asymmetry \citep{Lotz_2004}, and smoothness \citep{Lotz_2004} are  relevant for identifying potential clusters for follow-up observations; see \citet{Rasia_2013} and \citet{Ghirardini_2022} for descriptions of common morphology parameters. Morphologically accurate prediction images will allow observers to efficiently select clusters based on dynamical state and core type, which will not only improve our understanding of those key cluster properties, but also advance our understanding of physics more broadly. For example, selection of clusters by dynamical state is important for the study of dark matter \citep[e.g.,][]{Eckert_2022}.

In addition to morphology, we evaluate our model's capacity for de-noising observations and predicting the total flux of the observation. Like most forms of astronomical observation, X-ray observations suffer from foreground and background contamination. While both \textit{eROSITA}-like and \textit{Chandra}-like observation strategies suffer from these issues, \textit{eROSITA}'s shorter observation time and poorer angular resolution make it more challenging to disentangle signal from noise. Our follow-up merit assessment tool compensates for these differences by reducing background and differentiating between the emission from active galactic nuclei (AGN) and ICM.

Our tool uses a novel image-to-image convolutional neural network (CNN) trained on observations of simulated galaxy clusters from the hydrodynamic cosmological simulation \texttt{Magneticum} \citep{magneticum}. We accessed \texttt{Magneticum} via the Cosmological Web Portal \citep{cosmo_web_portal}, which simulates cluster mock observations with the \texttt{PHOX} algorithm \citep{PHOX1,PHOX2}. We used \texttt{SIXTE} to simulate realistic \textit{eROSITA} observations \citep{SIXTE}. A machine learning algorithm, which by its nature learns from inputted data, is appropriate for our task because it is the most capable at learning complicated nonlinear signals and correlations in data, as we would expect to exist between \textit{eROSITA}-like observations and the underlying astrophysical sources being observed \citep[see][for a review of deep learning]{Schmidhuber_2014}. We choose to use a CNN in particular, because of its capability to learn localized patterns in inputted data, like gradients, textures, and patterns. CNN's have been critical to advances in image-to-image prediction and analysis \citep[e.g.,][]{UNET,ploss}. CNN's have been applied to a variety of problems in astronomy like cosmic web simulations \citep{Rodriguez_2018}, exoplanet atmospheres \citep{Zingales_2018}, image reconstruction \citep{Flamary_2016}, and image denoising \citep{Vojtekova_2021}. In applying deep learning methods to galaxy cluster observations using cosmological simulations, we build on an established and proven practice \citep[see e.g.,][]{Ntampaka_2015,Green_2019,Ntampaka_2019}. Our work is also complementary to image-to-image neural network galaxy cluster cosmology work, like the SZ-effect image-emulator used in \citet{Rothschild_2022} or the XMM Newton X-ray super resolution and denoising algorithms developed in \citet{Sweere_2022}.

We present a machine learning approach to enhancing \textit{eROSITA} images to assess them for follow-up. This paper is organized as follows: In Section \ref{Methods} we describe simulated data that were used (\S\ref{data}), the structure of the algorithm (\S\ref{algorithm}), and the manner in which it was trained (\S\ref{training}). In Section \ref{results} we describe the performance of our model for each of the morphological parameters (Concentration \S\ref{concentration}, Asymmetry \S\ref{asymmetry}, Smoothness \S\ref{smoothness}), total flux (\S\ref{conserved flux}), background reduction (\S\ref{background reduction}), and mass dependence (\S\ref{Mass Dependence}). In Section \ref{discussion} we discuss the implications and limitations of our model. Section \ref{conclusion} is the conclusion.

\section{Methods}\label{Methods}
\begin{figure*}
    \centering
    \null \vspace{-20pt}
    \includegraphics[width=6in]{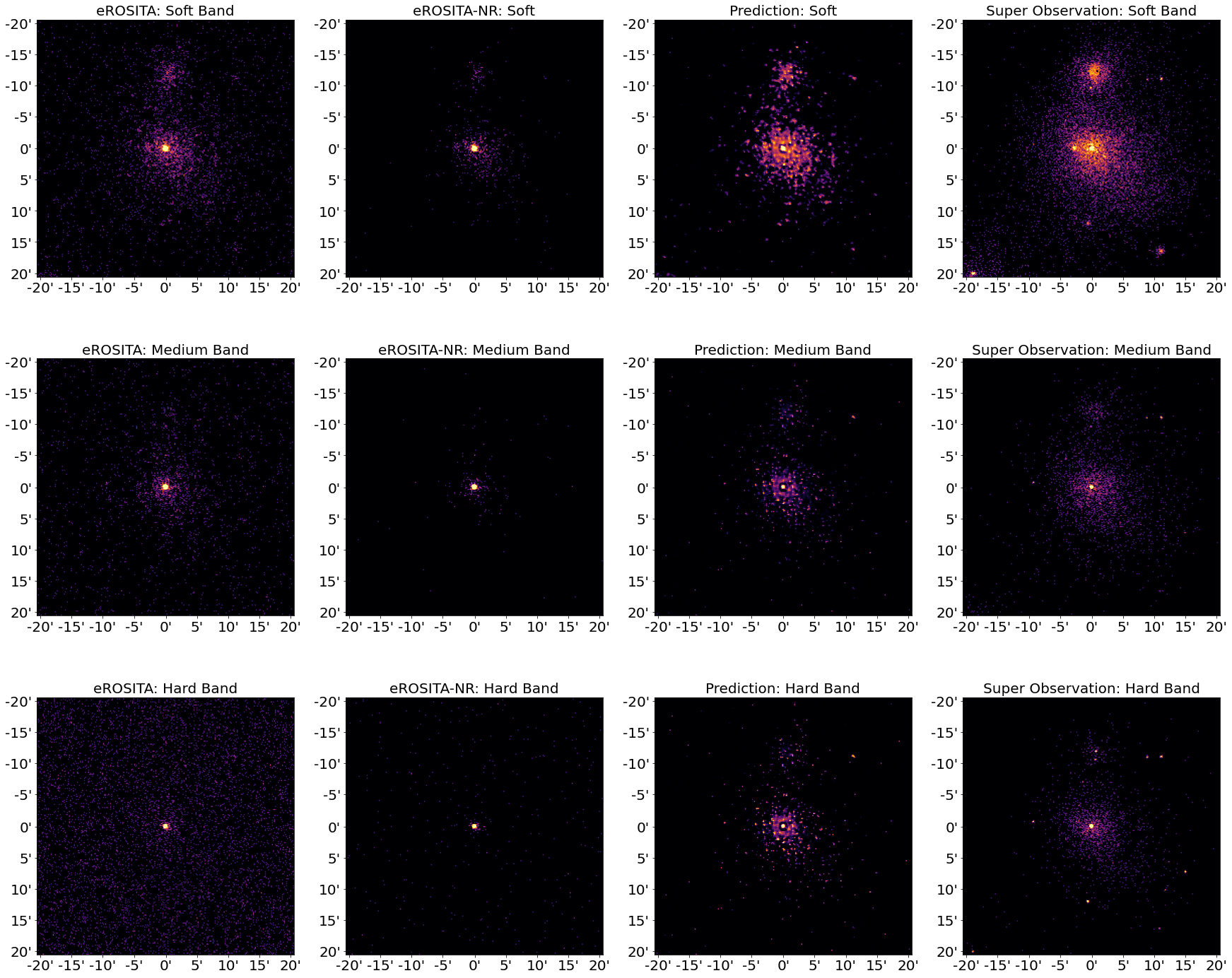}
    \caption{\raggedright Example observations of a sample galaxy cluster. The cluster shown has $M_{500c}=8.2\times10^{13}M_{\odot}$ and is at $z=0.07$. Rows correspond to the soft, medium, and hard X-ray energy bands. Columns correspond to different observation types (see \S\ref{data}), from left to right:  the \textit{eROSITA} mock image used as the machine learning input (``eROSITA''), a background subtracted image (``eROSITA-NR''), the machine learning output (``Prediction''), and the ground truth long duration mock observation (``Super Observation''). The flux color mapping is in log space, with the same color scaling for each image, from a color minimum at 0 photons per pixel to a color maximum of 10 photons per pixel. The predictions of the machine learning method developed in this research visibly outperform the \textit{eROSITA} and background-subtracted \textit{eROSITA} observations.  While the soft band \textit{eROSITA} observation captures most of the shape of the central cluster and the merging cluster, the substructure of central cluster and the neighboring AGN are not obviously distinguishable. The soft band prediction, on the other hand, shows most of this information clearly. This superior performance is most visible in the hard band, where \textit{eROSITA} is less sensitive and more noisy. The hard band \textit{eROSITA} observation displays the core of the central cluster in a field of background emission, whereas the prediction clearly shows the full shape of the central and merging cluster. For a more quantitative comparison of the results, see Section \ref{results}.}
    \label{fig:example_results}
\end{figure*}

\subsection{Data}\label{data}
Machine learning methods, in combination with hydrodynamic cosmological simulations, offer a powerful tool for galaxy cluster science. Machine learning methods, especially the convolutional neural network variant we develop in \S\ref{algorithm}, are exceptional at learning complicated patterns in multi-dimensional data. The methods we use require ``labeled'' data, meaning we need many realistic observations of galaxy clusters paired with matching, longer duration, background-free observations, which we refer to as ``super observations.'' Moreover, the choice of data determines the utility of the algorithm, meaning we need realistic observations of accurately simulated clusters. We choose to use hydrodynamic cosmological simulations because they easily provide pairs of simulated \textit{eROSITA} observations and super observations, while also closely matching the observed properties of AGN and the ICM \citep[see e.g.,][]{Hirschmann_2014, Rasia_2015, magneticum}. 

This research requires a hydrodynamic cosmological simulation large enough to generate a diverse sample of clusters and also high enough resolution to accurately model cluster substructures. To satisfy these constraints, we use the hydrodynamic cosmological simulation \texttt{Magneticum} \citep{magneticum}. Specifically, we use Box2/hr \citep{Hirschmann_2014}, a 352 Mpc/h sized  box with $2 \times 1584^3$ particles. It has a dark matter particle mass $M_{DM} = 6.9\times10^{8} M_{\odot}$, gas particle mass $M_{gas}=1.4\times10^{8}M_{\odot}$, and provides 6927 clusters at redshift $z=0.07$ with masses above $10^{13}M_{\odot}$, with a variety of mass accretion histories. The simulation uses cosmology constraints from \citet{Komatsu_2011}; i.e., total matter energy density, $\Omega_M = 0.272$ with 16.8\% baryons, a cosmological constant, $\Omega_{\Lambda} = 0.728$, Hubble constant, $H_0 = 70.4$, spectral index of the primordial power spectrum, $n_{s} = 0.963$, and an amplitude of matter fluctuations, $\sigma_8 = 0.809$.

Machine learning algorithms are inherently data driven, and therefore careful consideration must be given to data selection. To avoid biasing our model towards the more plentiful low mass, low redshift, clusters, we chose a roughly uniform mass and redshift distribution of clusters. We did so by subsampling the available low mass, low redshift clusters. Each cluster is included in the data set only once, from a unique line of sight. Our data set has 3285 galaxy cluster observations. Observations have a depth of 10 Mpc, and include emission from the galaxy cluster, nearby neighboring galaxy clusters, and nearby AGN. The emission from AGN is simulated as detailed in \citet{Biffi_2018}. Observed galaxy clusters have a mass range of $3.16 \times 10^{13}$ $M_{\odot}$ to $1.17 \times 10^{15}$ $M_{\odot}$ and a redshift range of 0.07 to 0.47.

The goal of our work is to create an algorithm capable of predicting a high quality observation from a lower quality observation. To do so, we must train the algorithm using pairs of observations. The galaxy cluster observations are therefore split in two categories, mock \textit{eROSITA} observations and super observations. The mock \textit{eROSITA} observations begin with a field of view and resolution matching that of \textit{eROSITA}, with a field of view of roughly 1 degree and a pixel size of 9.6 arcseconds. The observation time of these images is 2 ks. Expected background particle emission, instrument response, and point spread function are simulated using the \texttt{SIXTE} software \citep{SIXTE}. The super observations have the same detector area, field of view, and resolution as the \textit{eROSITA} observations, but are background-free, lack any instrument response or point spread function, and have an observation time of 10 ks. Because our over-arching goal is to accurately predict follow-up observations including line-of-sight AGN sources, AGN sources within 10 Mpc of the central galaxy cluster are included in our super observations. 

A background-subtracted \textit{eROSITA} observation set, which we refer to as \textit{eROSITA}-NR, is used as a baseline to compare prediction effectiveness to a non-machine learning method. The per-pixel background is defined as the mean of all nonzero pixels in an annulus with an inner radius of 140 pixels and a width of 5 pixels. This annulus range is sufficiently large to be exterior to all of the clusters' $R_{500c}$ radii in our data set. This is important, because we use the $R_{500c}$ radius, which is the radius in which the mass density of the cluster is 500 times the critical density of the Universe, and as a measure of the extent of the cluster. After the subtraction, all pixels with flux values less than zero are set to zero.

Each observation is divided into three energy bands, corresponding to soft X-rays (0.5-1.2 keV), medium X-rays (1.2-2.0 keV), and hard X-rays (2.0-7.0 keV). These bands were chosen following the definitions of the \textit{Chandra}/ACIS science and source detection energy bands\footnote{https://cxc.harvard.edu/csc/columns/ebands.html}, and also to take advantage of the different spectral behavior of AGN, ICM, and \textit{eROSITA} particle background. The frequency of the photons is known exactly for super observations, but for mock \textit{eROSITA} observations photons are sorted by their observed \textit{eROSITA}-defined channel number (PHA channel). The soft, medium, and hard X-rays are divided by the PHA bands 74-177, 178-274, and 275-722 respectively. Due to memory constraints, we only used the inner 256$\times$256 pixels of each image. This reduces the field of view to 40.96 arcminutes, but leaves the resolution unchanged. A summary of the observation image information is shown in Table \ref{tab:image_facts}. An example cluster, seen via mock \textit{eROSITA}, \textit{eROSITA}-NR, prediction, and super observations, is shown in Figure \ref{fig:example_results}.

\begin{table*}[h!]
\centering
 \begin{tabular}{||c| c c c c||} 
 \hline\hline
 Observation & Field of View & Pixel Size & Exposure Time & Background\\ [0.5ex]
  & (arcminutes) & (arcseconds) & (s)& Soft/Med/Hard
 (counts/pixel)\\ [0.5ex]
 \hline\hline
 \textit{eROSITA} &  40.96' & 9.6"$\times$9.6" & 2000& 0.02/0.02/0.1\\[1ex] 
 Super Observation &  40.96' & 9.6"$\times$9.6" & 10000& 0\\[1ex]
 \hline\hline
 \end{tabular}
 \caption{\label{tab:image_facts} A table of the image information. All observations types have the same field of view and the same angular resolution. They differ only in observing time and background level. Super observations are ideal observations; long exposure time, background-free, perfect instrument response, and no point spread function. Background counts are estimates from simulated blank sky observations that were put through our pipeline.}
\end{table*}

\subsection{Algorithm}\label{algorithm}

Convolutional neural networks are a class of machine learning algorithm that are often used for image processing tasks.  They make use of what are called ``convolutional filters.'' In the two dimensional case these can be understood as two dimensional matrices, where each element of the matrix is a parameter fitted during training. Images are convolved by sliding these filters over the image, taking the dot product of the matrix and a given region of the image. The size of these filter matrices and the method of sliding them over an image are hyperparameters determined by the user. During training these filters transition from random realizations to relevant feature detectors, commonly detecting edges, textures, and other patterns. By stacking layers of these filters on top of each other, the algorithm becomes capable of detecting more complex large scale features, like faces or animals. See \citet{Lecun_2015} for a review of deep learning and CNN's.

Our model is engineered to focus on accurately probing multiple length scales simultaneously. The images in our data set contain three primary components: ICM, AGN, and background, all of which are limited in size to length scales smaller than the entirety of the image. These small characteristic length scales encouraged us to abandon more traditional algorithm architectures, like the \texttt{UNET} \citep{UNET}, which rely on reducing images into a small set of globally important features. Our model instead consists primarily of two layers split into three paths. Each path has different sized filters (1$\times$1, 3$\times$3, 5$\times$5), each with 200 filters. We constructed and trained our model using the \texttt{Python} \citep{Python3} module \texttt{Tensorflow} \citep{tensorflow2015-whitepaper}.

\begin{figure*}
    \centering
    \includegraphics[width=.86\linewidth]{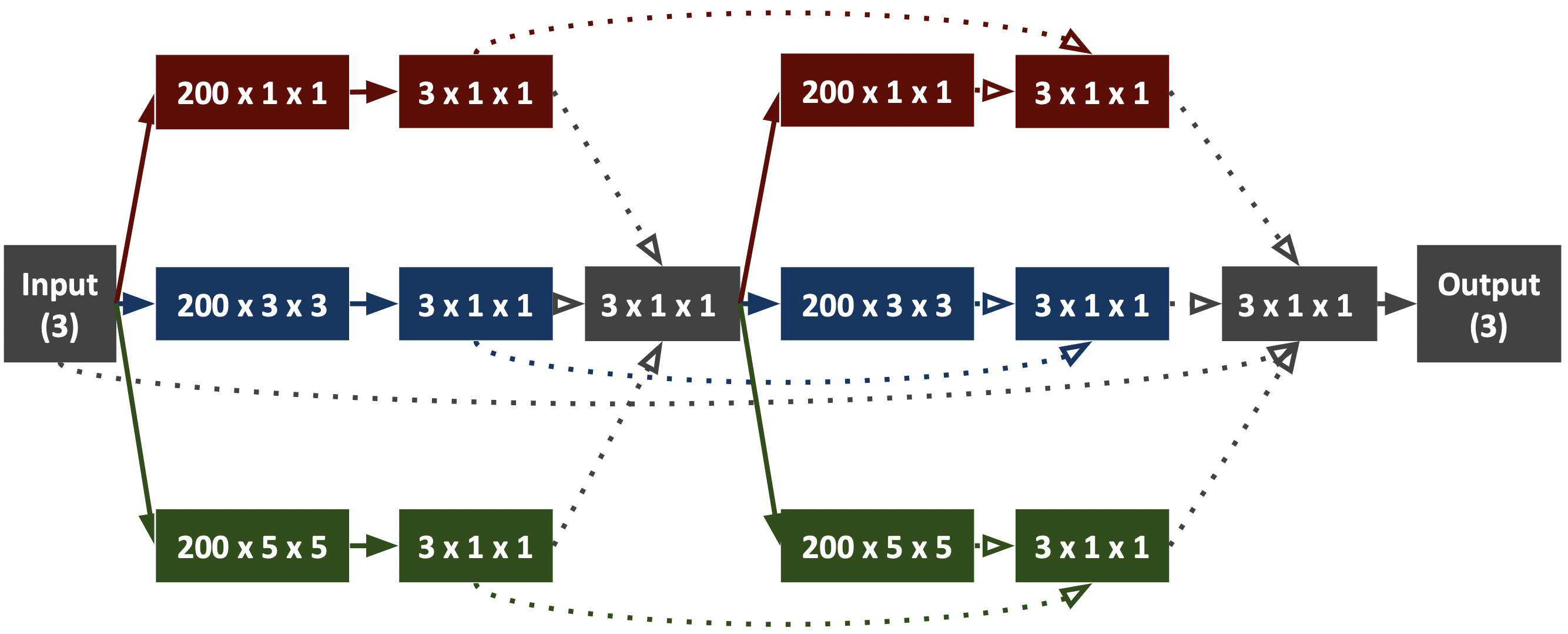}
    \caption{A diagram of the convolutional neural network architecture. We input a 256$\times$256 3-band pixel image. The three bands correspond to the three X-ray energy bands used, which were chosen to leverage both \textit{eROSITA}'s energy dependent sensitivity and the difference between AGN and ICM spectra. Further discussion of the energy bands is found in Section \ref{data}. The number of bands is shown in each box, either in parentheses or as the leftmost number. This is equal to the number of filters applied to the output of the preceding step (steps are linked by arrows). The three paths, each corresponding to a different filter size, are shown in different colors. Their filter sizes are the two rightmost numbers. From top to bottom the sizes of the filters in each path are 1$\times$1 (red), 3$\times$3 (blue), and 5$\times$5 (green). Dotted arrows indicate that a concatenation occurs before the filters are applied. This is done to probe different relevant length scales of an image without compressing the 256$\times$256 pixel size of the image. Instead, images are repeatedly reduced to 3-band images to reduce memory burden.}
\label{fig:Algorithm}
\end{figure*}

In the first layer, an input image observation is fed directly into each path. After having the filters applied to it, the image is compressed back into a 3-band image using 1$\times$1 filters, thus leaving us with three different 3-band images. These images are concatenated together to form a single 9-band image. This image is then also reduced into a 3-band image using 1$\times$1 filters. This composite 3-band image is then fed into another layer of the same structure as the first, albeit with differently trained weightings. The outputs of each of these paths in the second layer is then concatenated with the 3-band output of the corresponding path of the first layer. Filters are applied to these concatenations so that they are transformed into 3-band images. These three resulting 3-band images are then themselves, with the initial input, concatenated. The resulting 12-band image is then subsequently reduced to a 3-band image output. Our resulting model has 48,687 trainable parameters. A diagram of this algorithm is shown in Figure \ref{fig:Algorithm} and is described in Table \ref{tab:algorithm_breakdown}.

We chose these three different filter sizes to correspond to relevant image properties. Different astrophysical objects and features have different spatial and spectral properties, e.g. AGN flux is spatially compact and is brighter in hard X-ray while ICM flux is spatially correlated on larger scales and is dominant in soft X-ray. The 1$\times$1 filter path examines purely spectral information, comparing the ratio of the fluxes in each energy band, which we believe improves background suppression and AGN identification, as both of these image features have limited spatial correlations but unique spectral behaviors. The 3$\times$3 and 5$\times$5 filter paths then ought to identify spatial correlations from ICM, with two length scales chosen to account for both substructure and the varying pixel size of clusters given their mass distribution and redshift.

\begin{table*}
\centering
 \begin{tabular}{||c c c c c c||} 
 \hline
 Layer Name & Layer Type & Preceding Layer(s) & Filter \# & Kernel Size & Activation \\ [0.5ex] 
 \hline\hline
 x0 & Input & None & None & None & None \\ 
 x1 & Conv2D & x0 & 200 & 1$\times$1 & LeakyReLU\\
 x3 & Conv2D & x0 & 200 & 3$\times$3 & LeakyReLU\\
 x5 & Conv2D & x0 & 200 & 5$\times$5 & LeakyReLU\\
 x1b & Conv2DTranspose & x1 & 3 & 1$\times$1 & LeakyReLU\\ 
 x3b & Conv2DTranspose & x3 & 3 & 1$\times$1 & LeakyReLU\\
 x5b & Conv2DTranspose & x5 & 3 & 1$\times$1 & LeakyReLU\\
 x\_concat & Concat & x1b, x3b, x5b & None & None & None\\
 z0 & Conv2DTranspose & x\_concat & 3 & 1$\times$1 & LeakyReLU\\
 z1 & Conv2D & z0 & 200 & 1$\times$1 & LeakyReLU\\
 z3 & Conv2D & z0 & 200 & 3$\times$3 & LeakyReLU\\
 z5 & Conv2D & z0 & 200 & 5$\times$5 & LeakyReLU\\
 z1b & Concat & x1, z1 & None & None & None\\ 
 z3b & Concat & x3, z3 & None & None & None\\
 z5b & Concat & x5, z5 & None & None & None\\
 z1c & Conv2DTranspose & z1b & 3 & 1$\times$1 & LeakyReLU\\ 
 z3c & Conv2DTranspose & z3b & 3 & 1$\times$1 & LeakyReLU\\
 z5c & Conv2DTranspose & z5b & 3 & 1$\times$1 & LeakyReLU\\
 z\_concat & Concat & x0, z1c, z3c, z5c & None & None & None\\ 
 Output & Output (Conv2DTranspose) & z\_concat & 3 & 1$\times$1 & Linear\\[1ex] 
 \hline
 \end{tabular}
 \caption{\label{tab:algorithm_breakdown}A table of the convolutional neural network algorithm. A 256$\times$256 3-band pixel image is fed into the algorithm. Through an interconnected series of convolution and transpose convolution layers we achieve accurate predictions of cluster morphology while maintaining a smaller model. More details are available in Section \ref{algorithm} and Figure \ref{fig:Algorithm}.}
\end{table*}

\subsection{Training}\label{training}
Standard supervised machine learning training involves inputting data into an algorithm and then comparing the output (i.e., the prediction) to the label of the input (i.e., the truth value of the property of interest). The comparison is computed using a loss function. The weights of the algorithm are then changed to minimize the output of the loss function. In our case the inputs are the mock \textit{eROSITA} observations. The labels are the super observations. The loss function is a linear combination of the mean absolute error (i.e., the mean absolute difference between the pixels of the prediction image and the super observation)\footnote{We found that mean squared error performed poorly in the presence of background.} and ``morphology loss", defined as the linear combination of the mean absolute error of three morphology parameters; surface brightness concentration, asymmetry, and smoothness. We chose this loss function in order to emphasize the properties of the cluster we view as most important. By training minimizing the morphology loss of the algorithm, we improve the morphology parameters derived from predictions produced by the algorithm. We also tried a perceptual loss function, inspired by \citep{ploss}, using the third layer of the VGG19 network \citep{VGG19}. This is discussed in Section \ref{discussion}.

For the loss function we use ``fixed" versions of the morphology parameters, wherein either a fixed radius (in pixels) or the entire image is used to calculate the parameter. This is used in the algorithm because of its simplicity and consistency; it does not require information about the redshift or size of the central cluster. The fixed parameters are calculated as follows. The fixed surface brightness concentration is the ratio of the fluxes within 10 pixels and 100 pixels of the center of the image. The fixed asymmetry parameter is calculated using the absolute difference of the full image and the same image rotated 180 degrees, and is equal to the sum of the pixel values of this difference image normalized by the total flux of the original image. The fixed smoothness is calculated by applying an 11 pixel boxcar smoothing to the full original image, calculating the absolute difference between the smoothed image and the original image, summing the total flux of the difference image, and then normalizing that by the total flux of the original image. The fixed concentration, asymmetry, and smoothness parameters are described in equations \ref{fixed_c}, \ref{fixed_a}, and \ref{fixed_s}, respectively.  $\mathbf{X}$ is the observation image, $\mathbf{X_{180}}$ is the observation image rotated 180 degrees, $\mathbf{\tilde{X}}$ is the smoothed observation image, $F$ is the total flux with in some radius $r$ (if $r$ is unstated, the full image is used), where $r$ is in units of pixels. Examples of morphology parameter extremes, albeit for the $R_{500c}$ versions described in Section \ref{results}, are shown in Figure \ref{fig:morph_extremes}.

\begin{equation}\label{fixed_c}
    C = \frac{F(r \leq 10)}{F( r \leq 100)}
\end{equation}
\begin{equation}\label{fixed_a}
    A = \frac{F(|\mathbf{X}-\mathbf{X_{180}}|)}{F(\mathbf{X})}
\end{equation}
\begin{equation}\label{fixed_s}
    S = \frac{F(|\mathbf{X}-\mathbf{\tilde{X}}|)}{F(\mathbf{X})}
\end{equation}
Our training set, the data we set aside specifically for training the weights of the algorithm, constitutes 80\% of the full data set. 10\% of the remaining data forms our validation set. The validation set is used to evaluate the training progress in order to determine when to stop training. The algorithm is saved after each epoch it achieves a minimal validation loss. This procedure is repeated until the validation loss minimum is stable for over 100 epochs. The final 10\% of the data forms the test set, which is used to analyze the efficacy of the fully trained algorithm. The full data set is shuffled prior to partitioning.

\section{Results}\label{results}
Prediction algorithms like the one we are proposing must have clearly defined use cases. To illustrate the ways in which our model might benefit observers, we have defined several important predictive capabilities. These metrics are informed by the nature of the problem we seek to solve:  a follow-up observation of an initial \textit{eROSITA} observation will have higher signal to noise, a more physically accurate and more constrained luminosity profile, more visible substructure, and a more definite shape.  A prediction algorithm therefore must seek to increase the signal to noise in the input image, increase the brightness of astrophysical sources while differentiating between extended sources and point-spread-function-blurred AGN, and predict the location and brightness of real substructure. Similarly, the enhanced properties of the prediction must provide useful information for survey selection. That is why we chose to prioritize morphology parameters, which provide useful information about the properties of the cluster, especially those properties such as dynamical state, mass, and core type, which are of particular importance to galaxy cluster research. Note that while we prioritize morphologically accurate observations, we do not intend our prediction observations to be used for morphology measurements directly. Morphology is simply a metric we use to grade image accuracy (see Section \ref{discussion} for a discussion on assessing image accuracy and morphology).

\begin{figure*}
    \centering
    \null \vspace{-20pt}
    \includegraphics[width=6in]{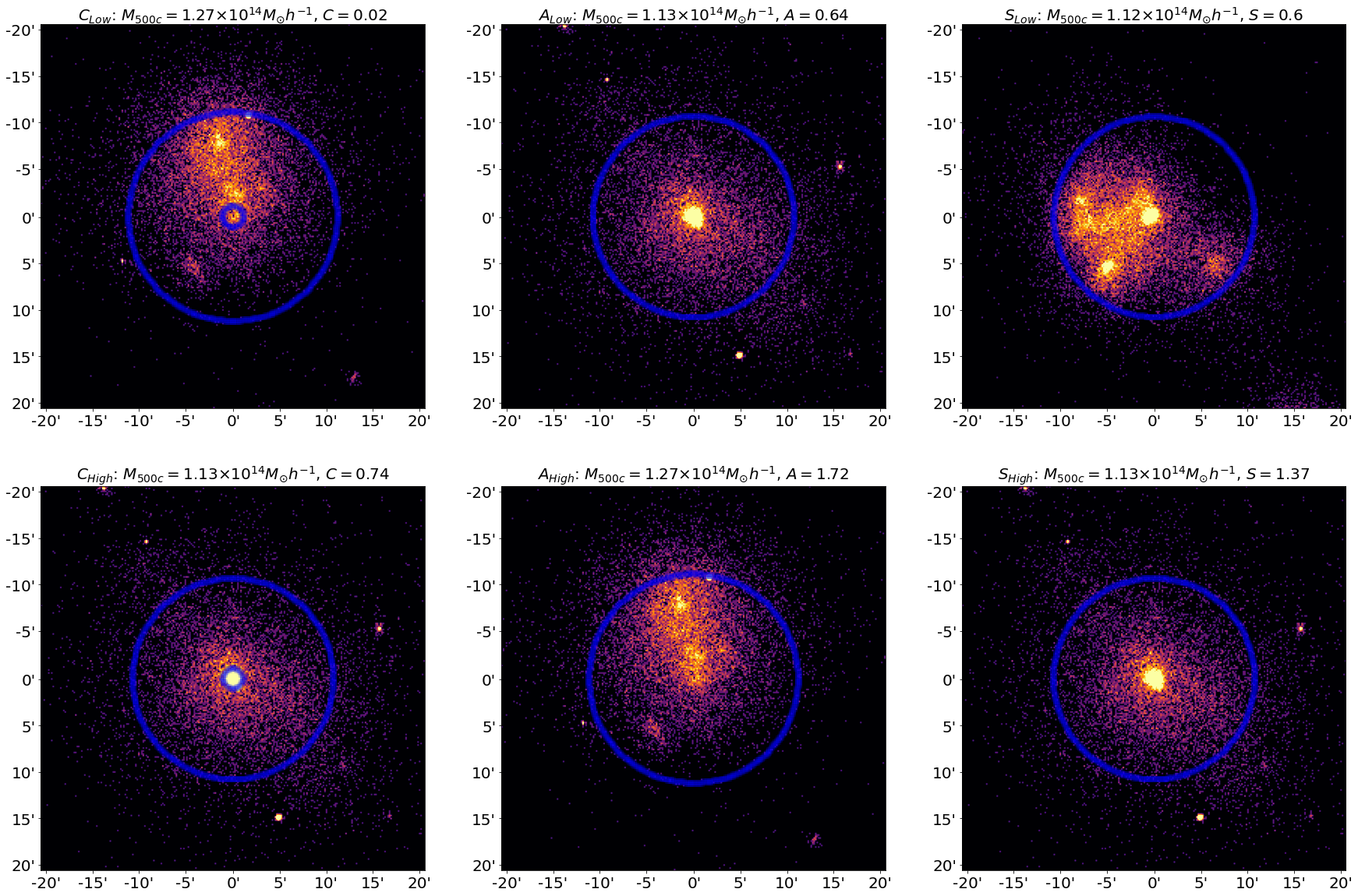}
    \caption{\raggedright Soft band observations of roughly equal mass clusters with varying $R_{500c}$ morphology parameters.  All clusters have a redshift of 0.07. Overlaid in blue is the $R_{500c}$ radius. For the leftmost column, the smaller blue circle is $.1\times R_{500c}$, which is used to calculate concentration.
    Left column:  An example of low (top) and high (bottom) concentration.
    Center column:  An example of low (top) and high (bottom) asymmetry.
    Right column:  An example of low valued (top) and high valued (bottom) smoothness. Larger values in the smoothness parameter indicate a peaky flux distribution. The bottom cluster is strongly concentrated (same cluster as bottom left) leading to higher smoothness value. The top cluster is clearly in a merger, which leads to a flatter flux distribution, meaning a more smoothly filled $R_{500c}$ radius.
    Note that the extremes in morphology parameter space tend to overlap, i.e., more centrally concentrated clusters tend to be more symmetric. For example, the same cluster is shown in the bottom left (highly concentrated) and top center (highly symmetric). Similarly, very asymmetric clusters tend to be less centrally concentrated. The top left cluster (less concentrated) and bottom center cluster (highly asymmetric) are the same.}
    \label{fig:morph_extremes}
\end{figure*}

Our intention is to aid in the selection of follow-up candidates, not to replace follow-up observations. Therefore, predictions must be probable enough to aid in discerning which clusters merit a follow-up. Metrics like those aforementioned provide an understanding for the accuracy of the predictions. An example prediction observation is shown in Figure \ref{fig:example_results}. Example observations of similar mass clusters with very different morphology parameters are shown in Figure \ref{fig:morph_extremes}. In the following subsections we examine the predicting power of our trained model for the three key morphology parameters (concentration \S\ref{concentration}, asymmetry \S\ref{asymmetry}, smoothness \S\ref{smoothness}), total flux (\S\ref{conserved flux}), and background reduction (\S\ref{background reduction}). We also evaluate the precision and accuracy of predictions as a function of mass (\S\ref{Mass Dependence}). We use the $R_{500c}$ versions of the morphology parameters. This formulation of the morphology parameters is more commonly used and more physically meaningful than the fixed version of the morphology parameters used in the loss function, but requires information about the cluster radius and redshift. Our true morphology parameters are calculated on observations with AGN, however the level of contamination from AGN in the soft band, where ICM dominates, is minimal.

\

\subsection{Concentration}\label{concentration}

Surface brightness concentration is a morphological parameter that estimates how centrally concentrated the mass of a galaxy cluster is, and is a key probe of a variety of cluster properties, including mass error \citep{Green_2019}, core type \citep{Santos_2008}, and dynamical state \citep{Rasia_2013, Parekh_2015, Lovisari_2017}. Given its ubiquity and usefulness as a metric, concentration is an important parameter for our prediction images to accurately replicate. Our $R_{500c}$ concentration parameter is modeled off the variant used by \citet{Lovisari_2017} and \citet{Green_2019}. The $R_{500c}$ surface brightness concentration calculates the ratio of the fluxes within $0.1 \times R_{500c}$ and within $R_{500c}$ of each cluster. Concentration varies from 0, a minimally concentrated cluster, to 1, a maximally concentrated cluster. Equation \ref{eqn_c}, shown below, describes the concentration calculation. Here $F$ denotes the total flux with in some pixel radius $r$.
\begin{equation}\label{eqn_c}
    C = \frac{F(r \leq .1 \times R_{500c})}{F( r \leq R_{500c})}
\end{equation}
True concentration is the concentration derived from the super observations.  We find that concentration values derived from the machine learning prediction observations are consistently closer to the true values, compared to concentrations derived from the \textit{eROSITA} or \textit{eROSITA}-NR images. In our test set data we find predicted $R_{500c}$ concentrations differed from the truth value by $(\Delta C_{\mathrm{soft}},\Delta C_{\mathrm{med}},\Delta C_{\mathrm{hard}}) = (0.02_{-0.04}^{+0.05}, 0.08_{-0.09}^{+0.1}, -0.03_{-0.1}^{+0.08})$, with the +/- values indicating the 84th and 16th percentile values in our test set results. Predicted soft band concentrations therefore have a roughly three times smaller median difference than the \textit{eROSITA} concentrations, with smaller scatter. This superiority is a reflection of the trained model's ability to simultaneously reduce background while boosting signal. This becomes especially apparent in the concentration parameters of the hard band, where high background dramatically degrades the accuracy. Here the median difference of prediction observation derived concentration is half that of the background subtracted \textit{eROSITA} concentration, the next most accurate concentration for that energy band, with half as much scatter. 

\begin{figure*}
    \centering
    \null \vspace{-20pt}
    \includegraphics[width=6in]{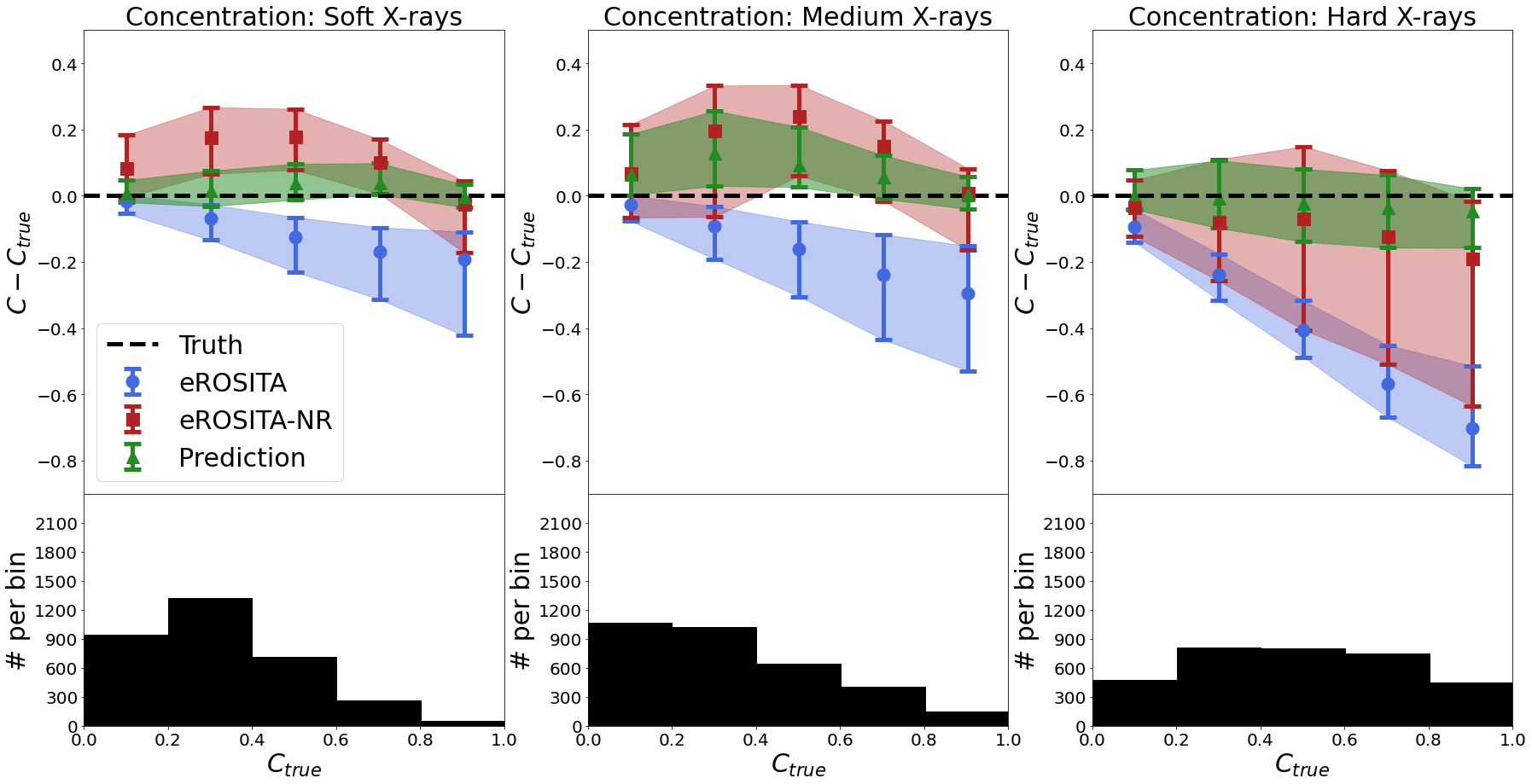}
    \caption{\raggedright $R_{500c}$ surface brightness concentration calculated for our full data set, as described in Section \ref{concentration}. The top plots show the difference between the calculated concentration from the \textit{eROSITA}, \textit{eROSITA}-NR (\textit{eROSITA} with background subtracted, see Section \ref{data}), and prediction observations each compared to the true concentration.  The true morphological parameters are calculated using the super observation. Results are binned by their true concentration value, with the number per bin shown in the histograms in the bottom panels. The median of the data in each bin (points) and the middle 68\% (error bars) are shown. For all energy bands, the prediction observation derived surface brightness concentrations are the most accurate. The median difference and scatter of those predictions appear to be independent of the true concentration, implying our prediction observations are valid for a diverse population of clusters.}
    \label{fig:concentration}
\end{figure*}

Plots of the concentration results as a function of the true (super observation) concentrations are shown in Figure \ref{fig:concentration}. In the top plots we show the median difference between the concentration derived by our prediction, the \textit{eROSITA} observation, or \textit{eROSITA}-NR observation and the super observation for the corresponding bins shown in the histograms directly below the plots. The shaded regions reflect the range of values in 68\% of the data in a given bin. The prediction-derived concentrations have a consistently lower bias and the results of the data have a consistently lower scatter across energy bands and true concentrations. Moreover, the bias and scatter of the concentration predictions are relatively consistent across true concentration bins, unlike the \textit{eROSITA}-derived results. Our results are evidence that a machine learning prediction model is better suited for accurately and precisely predicting concentrations of a diverse population of clusters than either the native \textit{eROSITA} observations or non-machine learning improvement. See Table \ref{tab:training_v_test} for a full quantitative comparison of the results.

\subsection{Asymmetry}\label{asymmetry}
The asymmetry of a cluster, meaning its deviation from circularity in its two-dimensional profile, is another key morphological parameter. Several metrics have been devised to probe asymmetry, including ellipticity and photon asymmetry \citep[see][]{Ghirardini_2022}. Our $R_{500c}$ asymmetry metric is modeled off of the variants used by \citet{Rasia_2013} and \citet{Green_2019}. We choose this formulation because of its simple implementation, usefulness as a probe of cluster dynamical state \citep{Rasia_2013}, and informativeness in cluster mass estimation \citep{Green_2019}. Symmetric clusters are less likely to be disturbed, and thus estimations of their mass are probably less biased. Asymmetry is an obvious selection criteria for a variety of cosmological uses, therefore accurately predicting asymmetry is necessary for our model to be useful.

The $R_{500c}$ asymmetry is calculated using the same procedure as the fixed asymmetry discussed in Section \ref{training}, but uses only pixels within $R_{500c}$. Asymmetry varies from 0, perfectly symmetric, to 2, maximally asymmetric. Equation \ref{eqn_a}, shown below, describes the calculation for asymmetry. Here $F$ denotes the total flux, $\mathbf{X}$ is the observation image, $\mathbf{X_{180}}$ is the observation image rotated 180 degrees, and $r$ is the radius within which the flux is calculated.

\begin{equation}\label{eqn_a}
    A = \frac{F(|\mathbf{X}-\mathbf{X_{180}}|;r\leq R_{500c})}{F(\mathbf{X};r\leq R_{500c})}
\end{equation}

\begin{figure*}
    \centering
    \null \vspace{-20pt}
    \includegraphics[width=6in]{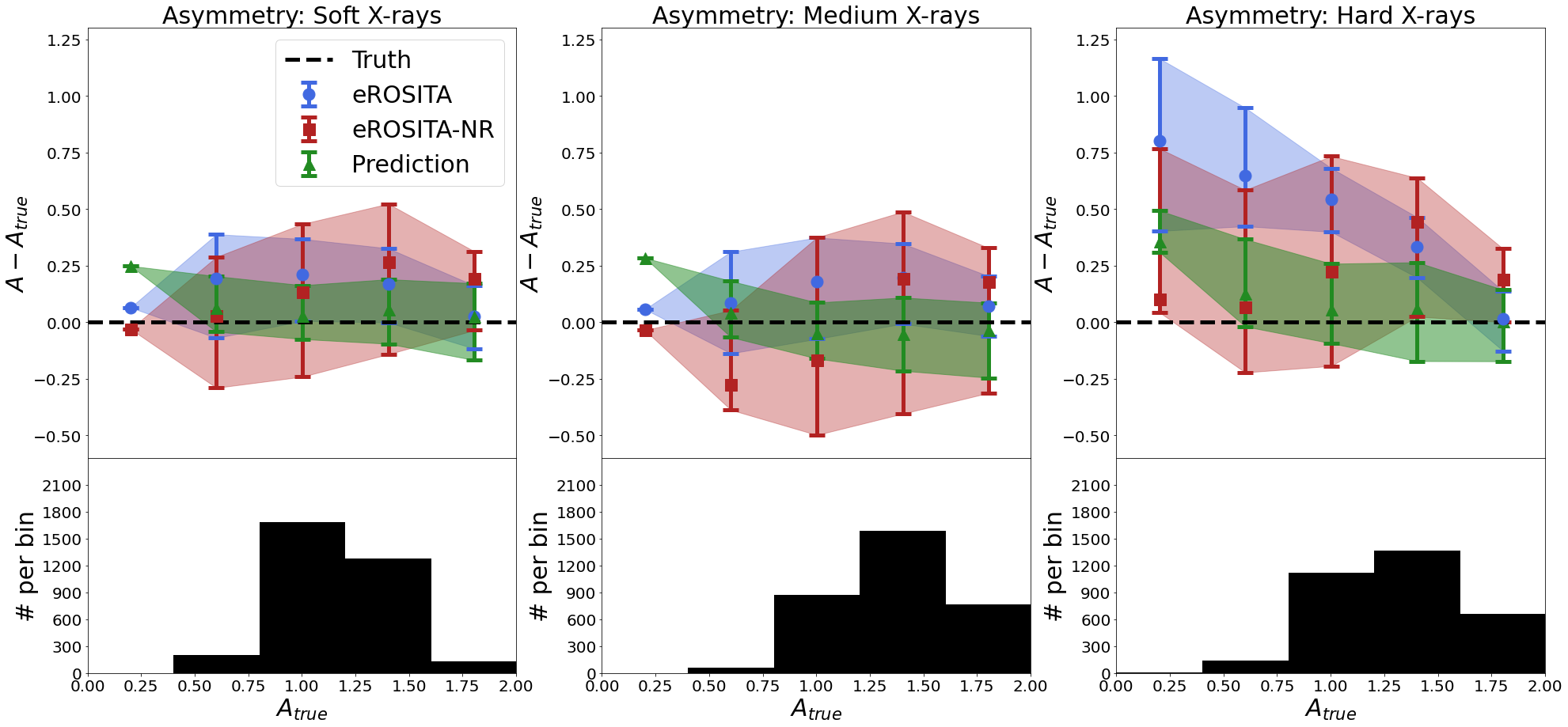}
    \caption{\raggedright Shown are plots of the $R_{500c}$ asymmetry calculated for our full data set, as described in Section \ref{asymmetry}. The top plots show the difference between the calculated asymmetry from the \textit{eROSITA}, \textit{eROSITA}-NR, and prediction observations and the true asymmetry, which is calculated using the super observation. Results are binned by their true asymmetry value, as shown in the histogram plots in the bottom row. The median of the data in each bin and the region representing 68\% of the results for the bins are plotted above. Prediction derived asymmetries are the most accurate, with lower median differences and much lower overall scatter. The median difference of asymmetry predictions appears to be dependent on true asymmetry, at least for low asymmetry clusters, suggesting the trained model overestimates the asymmetry of more circular clusters. The strength of this relationship is limited by the low number of highly symmetric clusters in our sample. Given the data-driven nature of machine learning algorithms, the limited number of highly symmetric clusters is probably the cause of the overestimation bias itself.}
    \label{fig:asymmetry}
\end{figure*}

Predicted asymmetry values are more closely correlated with super observation derived asymmetries than those from \textit{eROSITA} or \textit{eROSITA}-NR are. Asymmetry, compared to concentration, is poorly constrained in \textit{eROSITA} observations and is biased relative to super observation derived asymmetries. This is unsurprising, since asymmetry probes the less luminous outskirts of galaxy clusters, and is therefore more sensitive to background and short observation times. Despite this limitation, the prediction observation derived asymmetries are minimally biased across energy bands. In our test set data we found predicted $R_{500c}$ asymmetries differed from the truth value, as calculated using the super observations, by $(\Delta A_{\mathrm{soft}},\Delta A_{\mathrm{med}},\Delta A_{\mathrm{hard}}) = (0.03_{-0.12}^{+0.16}, -0.04_{-0.18}^{+0.16}, 0.06_{-0.24}^{+0.20})$. Like concentration, the strongest relative performance of the trained model's predictions for asymmetry is in the hard band. Here the median difference is $\sim6$ times smaller with comparable scatter compared to the \textit{eROSITA} results and $\sim4$ times smaller with nearly half as much scatter as the \textit{eROSITA}-NR results.  Plots of the asymmetry results as a function of the true (super observation) asymmetries are shown in Figure \ref{fig:asymmetry}. In the top plots we show the median difference between the super observation asymmetry and the asymmetry derived by our prediction, the \textit{eROSITA} observation, or \textit{eROSITA}-NR observation. The results are binned by the super observation derived asymmetry, with histograms of the bins shown directly below the plots. The shaded regions reflect the range of values in 68\% of the data in a given bin. Excluding the smallest asymmetry bin, which is too underpopulated to derive meaningful results, the bias in the asymmetry predictions is consistently lower than the other results. The asymmetry data set is not balanced across asymmetry parameter space, which is probably the cause of the slight dependence on true asymmetry. Very symmetric clusters are generally underrepresented in our data set, and therefore our model will struggle to accurately predict them. This could be corrected with access to more data. See Table \ref{tab:training_v_test} for more quantitative information about the results.

\subsection{Smoothness}\label{smoothness}
As they grow, galaxy clusters accrete nearby dark matter and baryons. This process results in clumpy substructure within the cluster, which can be visible in X-ray observations. The smoothness morphology parameter is a measure of the amount of substructure in a galaxy cluster, which in turn provides information about the dynamical state of the cluster \citep{Rasia_2013}. Like the other morphology parameters, smoothness also provides important information for cluster mass estimation \citep{Green_2019}. We adopt the smoothness parameter as defined in \citet{Green_2019}, which is similar to the fluctuation parameter used in \citet{Rasia_2013}.

To calculate the $R_{500c}$ smoothness, like the fixed smoothness discussed in section \ref{training}, we again apply an 11 pixel boxcar smoothing scale, but only to the pixels within $R_{500c}$ of the center. Equation \ref{eqn_s}, shown below, describes the calculation for smoothness. Here $F$ denotes the total flux, $\mathbf{X}$ is the observation image, $\mathbf{\tilde{X}}$ is the observation image after boxcar smoothing has been applied, and $r$ is the radius within which the flux is calculated.

\begin{equation}\label{eqn_s}
    S = \frac{F(|\mathbf{X}-\mathbf{\tilde{X}}|;r\leq R_{500c})}{F(\mathbf{X};r\leq R_{500c})}
\end{equation}

As is the case with all the key morphology parameters, predicted smoothness values are more closely correlated with super observation derived smoothness than those from \textit{eROSITA} or \textit{eROSITA}-NR are. In our test set data we found predicted $R_{500c}$ smoothnesses differed from the truth value, as calculated using the super observations, by $(\Delta S_{\mathrm{soft}},\Delta S_{\mathrm{med}},\Delta S_{\mathrm{hard}}) = (0.03_{-0.06}^{+0.08}, -0.02_{-0.14}^{+0.1}, -0.03_{-0.08}^{+0.11})$. See Table \ref{tab:training_v_test} for more information.

\begin{figure*}
    \centering
    \null \vspace{-20pt}
    \includegraphics[width=6in]{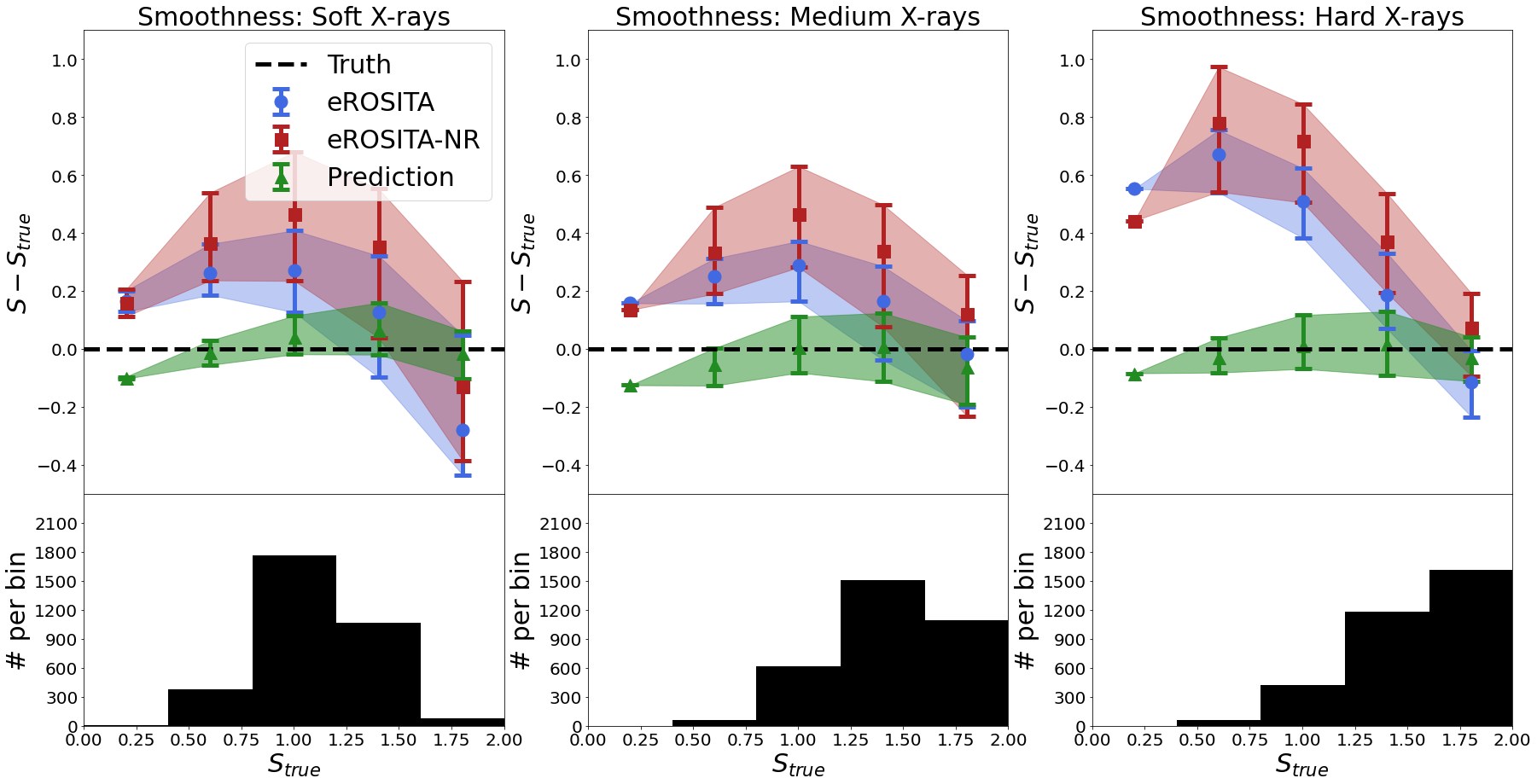}
    \caption{\raggedright Plots of the $R_{500c}$ smoothness calculated for our full data set, as described in Section \ref{smoothness}. The top plots show the difference between the calculated smoothness from the \textit{eROSITA}, \textit{eROSITA}-NR, and prediction observations compared to the true smoothness, which is calculated using the super observation. Results are binned by their true smoothness value, as shown in the histogram plots in the bottom panels. The median of the data in each bin and the region representing 68\% of the results for the bins are plotted above. Of the morphology parameters calculated, smoothness is the most biased by low observation time, which is why \textit{eROSITA}-derived smoothness values differ so greatly from the true values. Low observation time increases the perceived clumpiness of the ICM because of Poisson sampling. Removing background via simple subtraction only exacerbates this bias by reducing the number of pixels observed to have photons from the ICM. The prediction model, which utilizes information about the full image smoothness, more effectively removes background while preserving signal, which is why it so strongly outperforms the other observation types in terms of accuracy. As with asymmetry, the distribution of true smoothness parameters is not uniform over the possible space, so one must use this machine learning method with care. Predictions of very smooth clusters, and very clumpy soft band clusters, are more likely to biased.}
    \label{fig:smoothness}
\end{figure*}

Plots of the smoothness results as a function of the true (super observation) smoothness are shown in Figure \ref{fig:smoothness}. In the top plots we show the median difference between the smoothness derived by our prediction, the \textit{eROSITA} observation, or \textit{eROSITA}-NR observation and the super observation for the corresponding bins shown in the histograms directly below the plots. The shaded regions reflect the range of values in 68\% of the data in a given bin. The plots illustrate the prediction derived results low bias and low scatter in the data set across energy bands and smoothness parameter space. Like in the case of the asymmetry results, the strength of the results shown are limited by the unbalanced coverage of smoothness parameter space. Very smooth clusters, and very clumpy clusters in the soft X-ray, are not well represented. Machine learning is by nature data driven, so a lack of data in this parameter space could result in more inaccurate results.

\subsection{Total Flux}\label{conserved flux}

\begin{figure*}
    \centering
    \null \vspace{-20pt}
    \includegraphics[width=6in]{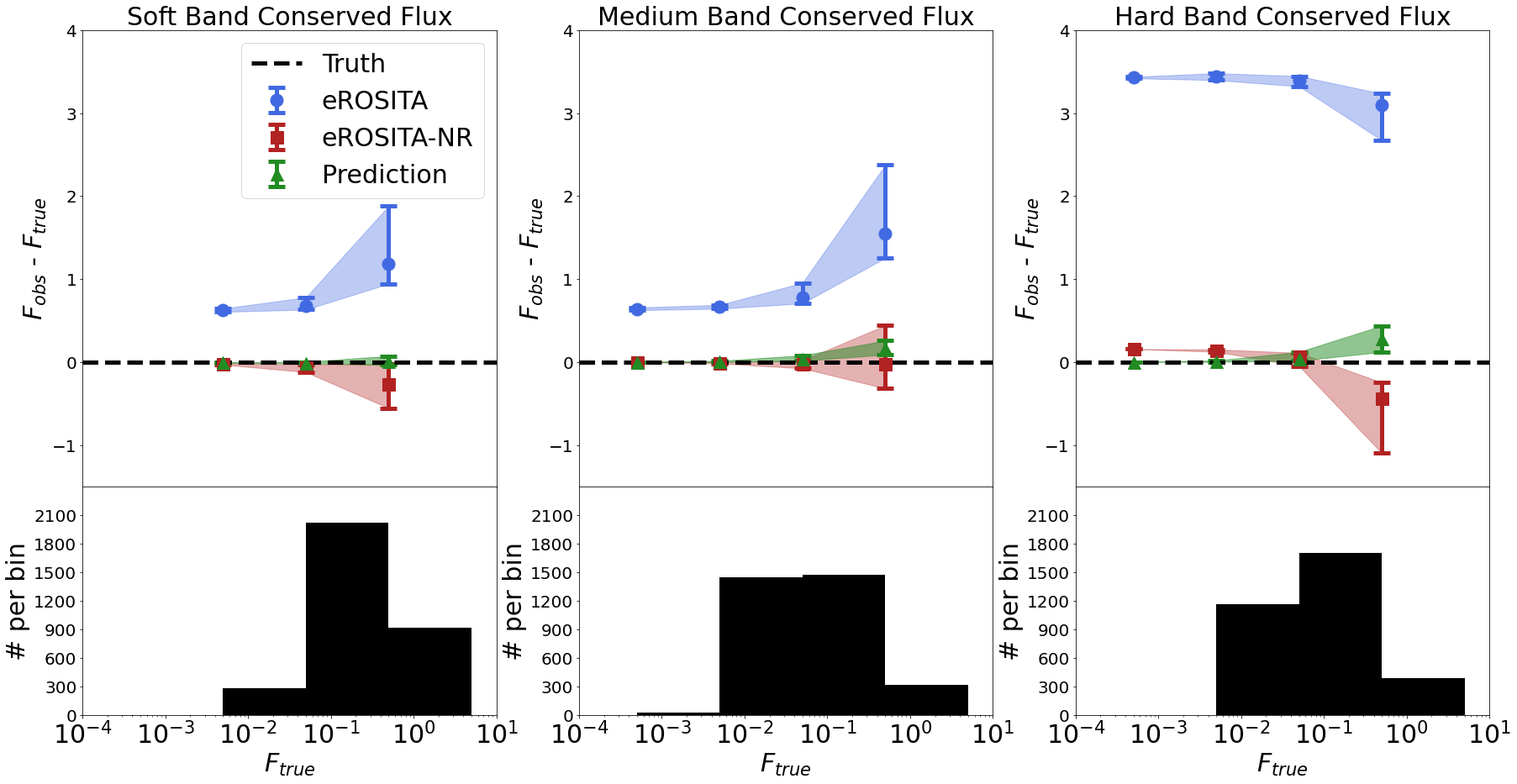}
    \caption{\raggedright Total flux for each observation. The format of the plots is the same as those of the morphology parameters. Total flux is defined as the total photon counts in an image normalized by the observing time. For the \textit{eROSITA} (plotted on the y-axis, shown in blue) and \textit{eROSITA}-NR  (plotted on the y-axis, shown in red) observations, this means a normalization of 2000 seconds. For the super observations (plotted on the x-axis), this is a 10,000 second normalization. The prediction observations (on the y-axis, shown in green) are also treated as 10,000 second observations. Results are binned by the true total flux, as determined by the super observation. A histogram of the bins used for the corresponding energy band is shown in each bottom panel. The prediction observations better preserve total flux than the \textit{eROSITA} or \textit{eROSITA}-NR observations. In the case of all of the \textit{eROSITA} observations and the \textit{eROSITA}-NR hard band observations this is because background photons dominate the image for dim sources. Even when \textit{eROSITA}-NR observations preserve the total flux decently well, as in the soft and medium energy bands, the prediction observations outperform them by maintaining a lower scatter.\newline}
    \label{fig:flux}
\end{figure*}

Do prediction observations conserve the total flux of the observation? Because morphology parameters all involve normalization by the total flux, they shed little light on that question. Conservation of the total flux is a potentially important feature for prediction observations, however, and merits investigating. We perform a simple test of this and find that the prediction observations do preserve the flux, and do so with less scatter than the background-subtracted \textit{eROSITA} observations. See Figure \ref{fig:flux} for a plot of the results. Figure \ref{fig:flux} illustrates the power of our model in conserving total flux, especially with regards to the soft band, where the scatter in the difference between predicted and true total flux is negligible. The results are also listed in Table \ref{tab:training_v_test}.

\begin{table*}[h!]
\centering
 \begin{tabular}{||c c c c c c c||} 
 \hline\hline
  &  &  &  &  & & \\
 Observation & Band & Concentration & Asymmetry & Smoothness & Total Flux & Background\\ [0.5ex] 
 \hline\hline
 Training + Validation &  &  &  &  & & \\
 \hline\hline
 \textit{eROSITA} & Soft & $-0.07_{-0.1}^{+0.05}$ & $0.19_{-0.19}^{+0.17}$ & $0.23_{-0.2}^{+0.15}$ & $0.72_{-0.08}^{+0.45}$& $1115_{-369}^{+74}$ \\[1ex] 
 \textit{eROSITA}-NR & Soft & $0.14_{-0.12}^{+0.1}$ & $0.18_{-0.4}^{+0.29}$ & $0.41_{-0.25}^{+0.22}$ & $-0.06_{-0.18}^{+0.04}$& $10_{-4}^{+4}$ \\[1ex]
 Prediction & Soft & $0.02_{-0.04}^{+0.06}$ & $0.04_{-0.12}^{+0.13}$ & $0.04_{-0.07}^{+0.09}$ & $-0.06_{-0.26}^{+0.05}$& $138_{-37}^{+19}$ \\[1ex]
 \hline
 \textit{eROSITA} & Medium & $-0.09_{-0.15}^{+0.07}$ & $0.16_{-0.21}^{+0.17}$ & $0.12_{-0.2}^{+0.17}$ & $0.71_{-0.05}^{+0.27}$& $1222_{-179}^{+54}$ \\[1ex] 
 \textit{eROSITA}-NR & Medium & $0.14_{-0.2}^{+0.15}$ & $0.12_{-0.56}^{+0.31}$ & $0.27_{-0.25}^{+0.22}$ & $-0.01_{-0.04}^{+0.02}$& $11_{-3}^{+4}$ \\[1ex]
 Prediction & Medium & $0.08_{-0.08}^{+0.13}$ & $-0.04_{-0.15}^{+0.14}$ & $-0.02_{-0.12}^{+0.11}$ & $0.0_{-0.05}^{+0.01}$& $100_{-15}^{+15}$ \\[1ex]
 \hline
 \textit{eROSITA} & Hard & $-0.38_{-0.25}^{+0.22}$ & $0.37_{-0.3}^{+0.21}$ & $0.04_{-0.21}^{+0.33}$ & $3.41_{-0.11}^{+0.06}$& $6530_{-789}^{+193}$ \\[1ex] 
 \textit{eROSITA}-NR & Hard & $-0.09_{-0.27}^{+0.18}$ & $0.29_{-0.37}^{+0.33}$ & $0.22_{-0.22}^{+0.34}$ & $0.1_{-0.21}^{+0.04}$& $311_{-39}^{+22}$ \\[1ex]
 Prediction & Hard & $-0.02_{-0.11}^{+0.09}$ & $0.05_{-0.17}^{+0.18}$ & $-0.01_{-0.09}^{+0.1}$ & $0.01_{-0.04}^{+0.01}$& $152_{-24}^{+27}$ \\[1ex]
 \hline\hline
 Test Data  &  &  &  &  &  &\\
 \hline\hline
  \textit{eROSITA} & Soft & $-0.07_{-0.09}^{+0.05}$ & $0.17_{-0.2}^{+0.17}$& $0.22_{-0.22}^{+0.15}$ & $0.72_{-0.08}^{+0.49}$& $1107_{-348}^{+78}$ \\[1ex] 
 \textit{eROSITA}-NR & Soft & $0.14_{-0.1}^{+0.1}$ & $0.16_{-0.39}^{+0.32}$ & $0.39_{-0.26}^{+0.24}$ & $-0.07_{-0.18}^{+0.04}$& $10_{-4}^{+4}$ \\[1ex]
 Prediction & Soft & $0.02_{-0.04}^{+0.05}$ & $0.03_{-0.12}^{+0.16}$ & $0.03_{-0.06}^{+0.08}$ & $-0.07_{-0.27}^{+0.05}$& $137_{-36}^{+19}$ \\[1ex]
 \hline
 \textit{eROSITA} & Medium & $-0.09_{-0.14}^{+0.07}$ & $0.17_{-0.22}^{+0.17}$ & $0.12_{-0.21}^{+0.17}$ & $0.71_{-0.06}^{+0.29}$& $1216_{-194}^{+58}$ \\[1ex] 
 \textit{eROSITA}-NR & Medium & $0.14_{-0.21}^{+0.15}$ & $0.13_{-0.58}^{+0.29}$ & $0.25_{-0.24}^{+0.27}$ & $-0.01_{-0.05}^{+0.02}$& $12_{-4}^{+3}$ \\[1ex]
 Prediction & Medium & $0.08_{-0.09}^{+0.1}$ & $-0.04_{-0.18}^{+0.16}$ & $-0.02_{-0.14}^{+0.1}$ & $0.0_{-0.05}^{+0.01}$& $100_{-15}^{+15}$ \\[1ex]
 \hline
 \textit{eROSITA} & Hard & $-0.37_{-0.26}^{+0.2}$ & $0.37_{-0.3}^{+0.23}$ & $0.02_{-0.2}^{+0.39}$ & $3.4_{-0.13}^{+0.06}$& $6519_{-802}^{+209}$\\[1ex] 
 \textit{eROSITA}-NR & Hard & $-0.07_{-0.23}^{+0.15}$ & $0.26_{-0.34}^{+0.38}$ & $0.21_{-0.24}^{+0.39}$ & $0.09_{-0.23}^{+0.05}$& $307_{-37}^{+24}$ \\[1ex]
 Prediction & Hard & $-0.03_{-0.1}^{+0.08}$ & $0.06_{-0.24}^{+0.2}$ & $-0.03_{-0.08}^{+0.11}$ & $0.0_{-0.06}^{+0.01}$& $151_{-19}^{+24}$ \\[1ex]
 \hline
 \end{tabular}
 \caption{\label{tab:training_v_test} A table of the median and 68\% intervals for the difference between the \textit{eROSITA}/\textit{eROSITA}-NR/Prediction derived values and the super observation derived values for morphology parameters (concentration, asymmetry, smoothness), total flux, and background removal. See Section \ref{results} for a description of each evaluation metric. See Section \ref{data} for information about the observation types and energy bands. Results for both the training-validation combined data set and test set are shown. Prediction derived results for the test set are consistent with those from the training-validation combined data set, suggesting the trained model has not over-fit the training data.}
\end{table*}

\subsection{Background Reduction}\label{background reduction}
Observations of galaxy clusters, like all astronomical observations, are hindered by various forms of background. Background in an observation has multiple sources. Background is generated both by non-galaxy cluster X-ray sources, particles, and as a product of the instrument itself. One potential use of an observation prediction model is to reduce background wherever possible. The machine learning model can achieve this by utilizing multi-wavelength information, as the cluster signal-to-background ratio of \textit{eROSITA} is higher at lower frequencies. Additionally, as the model learns the general shape of galaxy clusters and AGN it can better suppress pixels that do not conform to their luminosity profiles. It is important to reiterate that what the model produces is only a prediction and some background reduction will be inaccurate.

\begin{figure*}
    \centering
    \null \vspace{-20pt}
    \includegraphics[width=6in]{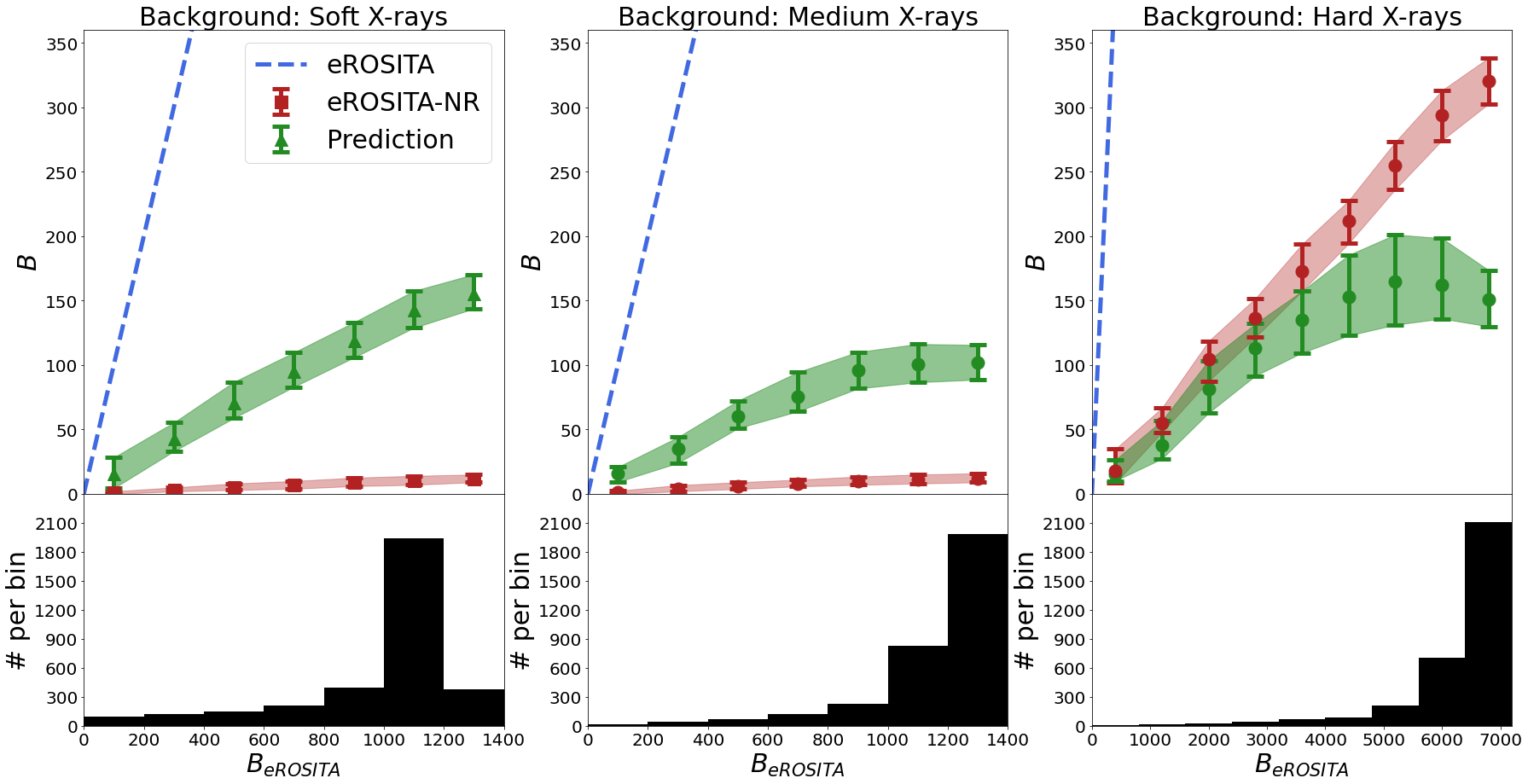}
    \caption{\raggedright Background calculated for our full data set, as described in Section \ref{background reduction}. The top plots show the background ($B$) from the \textit{eROSITA}, \textit{eROSITA}-NR, and prediction observations plotted against the \textit{eROSITA} observations background ($B_{eROSITA}$). Results are binned by their \textit{eROSITA} background, as shown in the histograms plots on the bottom. The median of the data in each bin and the region representing 68\% of the results for the bins are plotted above. The background subtraction method, described in Section \ref{data}, outperforms the prediction in background reduction for the soft and medium energy bands. The prediction observations have lower background in the hard band, however, likely due to the weak signal in \textit{eROSITA} observations at energies above 2 keV. Moreover, the prediction observations have the benefit of consistent background across energy bands, with only a weak dependence on the input observation's background. It should also be noted that the background subtraction method, while better at reducing background, also removes signal, and has worse morphology parameter accuracy.}
    \label{fig:noise}
\end{figure*}

We can quantify background reduction by comparing our background-free super observations to our \textit{eROSITA}-like input observations and the prediction observations. We apply a Gaussian smoothing filter to the super observations in order to mimic the effects of both the point-spread function and the difference in the random realization of photons between observations. We then catalogue the pixels with zero flux in the smoothed images. For the rest of our observations, we define the background as the total flux of those pixels. Plots of the background results are shown in Figure \ref{fig:noise}. Table \ref{tab:training_v_test} quantifies the results. 
We find that a background subtraction, as described in Section \ref{data}, outperforms our model's background suppression in the soft and medium energy bands. It does so at a cost to the accuracy in the $R_{500c}$ morphology parameters, especially asymmetry and smoothness (see Figures \ref{fig:asymmetry} and \ref{fig:smoothness}). The model outperforms subtraction in the hard energy band, however, where \textit{eROSITA} observations are especially background dominated. This advantage in performance could be useful for observations that are reliant on the hard band, like AGN-focused observations. Note that our mock \textit{eROSITA} observations only included simulated particle background, not X-ray foreground, or other X-ray background sources.

\subsection{Mass Dependence}\label{Mass Dependence}
\begin{figure*}
    \centering
    \null \vspace{-20pt}
    \includegraphics[width=6in]{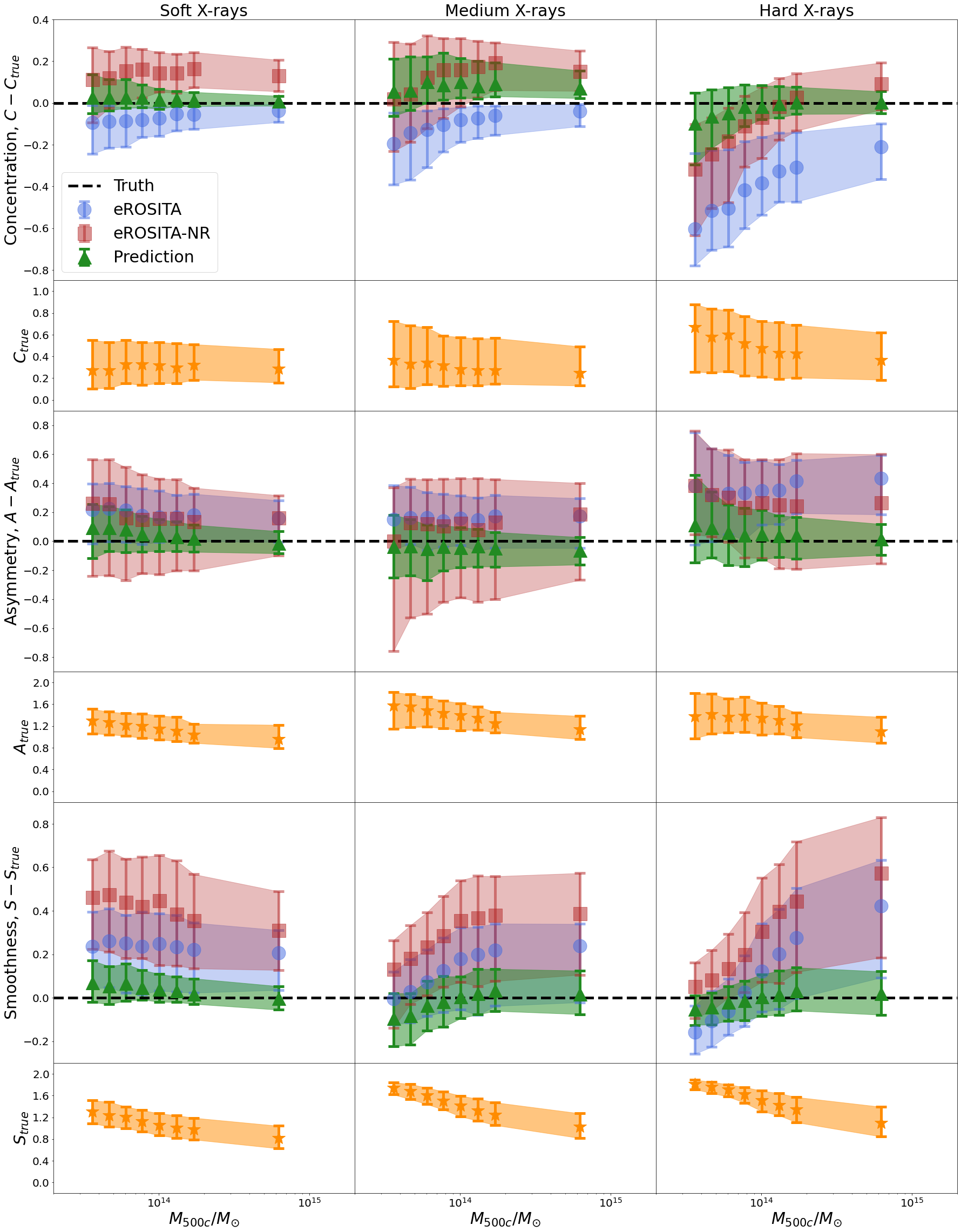}
    \caption{\raggedright Morphology parameter results are shown above. Morphology parameters, as derived from mock-\textit{eROSITA}, \textit{eROSITA}-NR (background subtracted), and prediction observations are compared to the true values, derived from the super observations. The top multicolored plots show the difference between observed parameter and true parameter, binned by mass. The bins are roughly equal in population. The points and error bars show the median and 16th to 84th percentile range of the results and are plotted in the middle of the mass bin range. The bottom orange plots illustrate the true parameter as a function of mass. The prediction derived constraints, shown in green, with the median values marked by triangles, are clearly superior to the alternatives. This is especially apparent in the soft band, where the median difference and scatter are small for all parameters. These results suggest our model produces morphologically accurate and precise observations without a noticeable mass bias for clusters between $\sim 10^{13}$ to $10^{15} M_{\odot}$.}
    \label{fig:mass_dep}
\end{figure*}

In addition to checking for bias in our prediction observations' morphological accuracy as a function of true morphology, we also examined whether there was any trend as a function of cluster mass. We found no obvious mass dependence in the soft band, which is the band most useful for measuring cluster morphology. The trends apparent in higher energy bands were very small, and the scatter in the difference between true and predicted parameters was roughly consistent across energy bands. These results suggest that the model predicts morphologically accurate and precise observations for clusters within our data sets mass range (i.e., $\sim 10^{13}$ to $10^{15} M_{\odot}$). See Figure \ref{fig:mass_dep} for a visualization of the results.

\section{Discussion}\label{discussion}
Our model demonstrates the powerful potential of machine learning in assessing the merit of follow-up observations. We have shown that our prediction images preserve morphology, which correlates to important galaxy cluster properties like dynamical state and core type. This is important, because to resolve outstanding questions about these properties we need high resolution, long duration, observations from instruments like \textit{Chandra}, which have limited availability. The model's useful prediction observations will allow galaxy cluster observers to effectively and efficiently evaluate which \textit{eROSITA}-observed clusters merit follow-up observation. Predicted observations are not a replacement for follow-up observations, nor are predicted morphology parameters a replacement for real morphology parameter measurements, but predictions are useful for determining if an \textit{eROSITA}-observed cluster is likely to have a property of interest (e.g., a cool core) that merits a follow-up observation. Machine learning models are most effective when used in an ethical way, with an understanding of their strengths and limitations. There are ways in which an informed user might adapt or improve an model. Below, we discuss the strengths and limitations of our model and the ways in which one might overcome those limitations.

\subsection{Classification}\label{classification}
A strength of our model, as we have shown, is that our prediction observations are more morphologically accurate than either \textit{eROSITA} or \textit{eROSITA}-NR observations. It is worth noting, however, that this relative accuracy can be applied in a variety of unexpected ways. For example, imagine an observer wishes to perform follow-up observations on only the most highly disturbed clusters. To do so, they would like to determine which clusters are in the 90th percentile of asymmetry. We can test this premise on our own data, by examining whether the 10\% most asymmetric clusters according to the prediction observations match the 10\% most asymmetric clusters as determined by the super observations. In doing so, we find that our trained model can determine whether a cluster is among the 10\% most asymmetric clusters with a true positive rate of 59\% and a false positive rate of 5\%. In other words, 59\% of clusters in the 90th percentile of true asymmetries are found in the 90th percentile of predicted asymmetries. Of clusters that are not in the 90th percentile of true asymmetries, only 5\% are found in the 90th percentile of predicted asymmetries. For identifying the 90th percentile of soft band concentration or soft band smoothness, the true positive (TPR) and false positive (FPR) rates improve to $\rm{TPR}_{C} = 82\%$, $\rm{FPR}_{C} = 2\%$, $\rm{TPR}_{S} = 73\%$, and $\rm{FPR}_{S} = 3\%$. The number of possible classifications is infinite and we cannot provide an exhaustive list of the true positive and false positive rates, but we encourage users to recognize this potential application of our model.

\subsection{Domain Shift}\label{Domain Shift}
A drawback of using simulated data is a problem known as domain or distributional shift \citep[see section 7 of ][for a discussion]{Amodei_2016}. Simulated data will invariably differ from real data. These differences, rooted in both the computational constraints of simulating a universe and in our still limited understanding of important cosmological and astrophysical phenomena, will limit the utility of a model trained solely on simulated data. The morphology parameters, by their nature, are deeply intertwined with the baryonic physics of galaxy clusters \citep[e.g.,][]{Lau_2011, Lau_2012, Chen_2019, Fernando_2021}. These physics are the most difficult to simulate, and the least well understood to model, and therefore often vary from cosmological simulation to cosmological simulation. This increases the likelihood that biases induced by domain shift will be present when our model is applied to real data, or data from a different cosmological simulation. In addition to simulation related biases, we are limited by the morphological parameter space of our data set. Clusters at the extremes of the parameter space, e.g., very smooth clusters, are relatively unrepresented. In the regions of cluster morphology parameter space with limited training data (see the bottom plots of Figures \ref{fig:concentration}, \ref{fig:asymmetry}, and \ref{fig:smoothness}) one should use caution with the assessment tool's predictions. 

A potential solution to this problem is to make use of what is known as transfer learning. We can retrain a model, previously trained on simulated observations, on pairs of observations (e.g., \textit{eROSITA} and \textit{Chandra} observations of the same clusters). This technique can correct the distributional shift between different data sets, in this case simulated and real data. It has the additional advantage of requiring many fewer real data observation pairs than would be required to train a new model completely on real data. This technique has been applied to a variety of problems in astronomy \citep[e.g.,][]{Ackermann_2018, Dominguez_2019, Perez_Carrasco_2019}.

\subsection{Redshift, Observing Time, and Resolution Dependence}\label{biases}
\begin{figure*}
    \centering
    \null \vspace{-20pt}
    \includegraphics[width=6in]{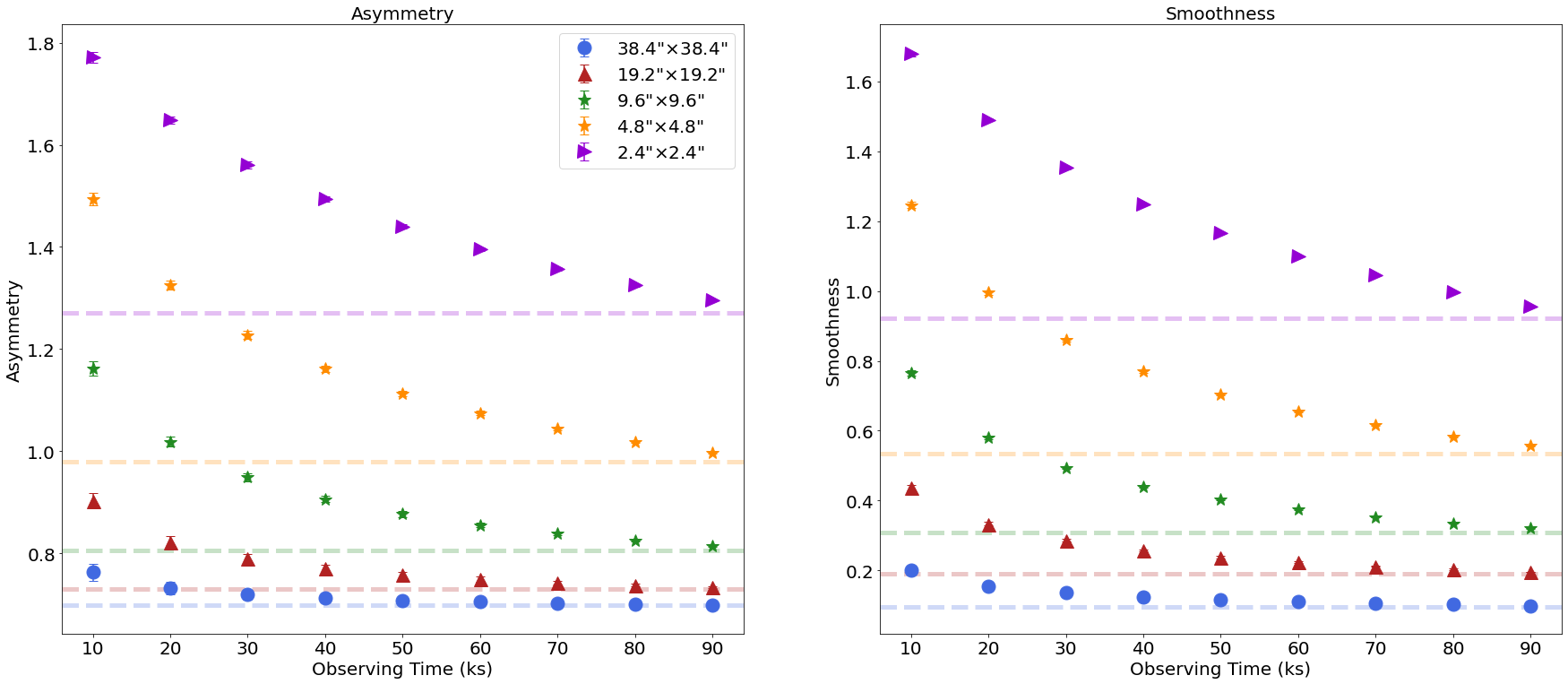}
    \caption{\raggedright Shown above are plots of the observed asymmetry and smoothness of a single cluster as a function of observing time and resolution. A single 100,000 second background-free mock observation of a $M_{500c}=1.48\times10^{14}M_{\odot}h^{-1}$ cluster at redshift $z=0.07$ with \textit{eROSITA} field of view was taken from \texttt{Magneticum} using \texttt{PHOX} \citep{magneticum,PHOX1,PHOX2}. Observing times were simulated by sampling different sized fractions of the photon list. The different colored and shaped points illustrate different pixel resolutions, denoted by the pixel size in arcseconds. The images we used for training our model, like \textit{eROSITA}, have a pixel size of 9.6"$\times$9.6". Error bars, representing the scatter of 1000 bootstrap samples, are also shown, but, in most cases, are smaller than the points. The dashed lines illustrate the parameter values for the full 100 ks observation. This plot illustrates that morphology parameters change depending on observing time and resolution, but also suggests that intrinsic uncertainty of morphology parameters on a 10 ks observation are small, albeit dependent on the intrinsic brightness of the source. Concentration is not shown, as its variation as a function of resolution and observing time is negligible.}
    \label{fig:morphology_variance}
\end{figure*}

We are not the first to recognize the importance of morphology in sample selection.  For example, \citet{Mantz_2015} developed the symmetry–peakiness–alignment morphology specifically to provide a more robust characterization of cool core clusters across different redshifts and observing instruments. While we opt for an alternative construction of morphology for computational reasons, our concentration-asymmetry-smoothness morphology provides similar information about cool cores, and we characterize robustness issues related to observing instrument and redshift. In the case of redshift, we find no substantial redshift bias in concentration, asymmetry, or smoothness. This is true across all energy bands, but especially in the soft band. We also performed a minor analysis of the effects of different observing instruments ourselves, investigating the change in morphology parameter value for an individual cluster as a function of resolution and observing time (see Figure \ref{fig:morphology_variance}). We find parameter values do change as a function of observing time and resolution. However, our results also suggest that the uncertainty of morphology parameters is minimal, even for lower observing times and coarser resolutions. This suggests that, given a representative sample of clusters observed by a single instrument, the morphology parameters of a cluster still provide useful information when compared to the distribution of morphology parameters of clusters observed by that instrument. Given the number of clusters \textit{eROSITA} is expected to observe, this constraint is not problematic. Moreover, while morphology parameters derived from the observations of one instrument do not directly map to those derived from the observations of another, the relationship between morphology and observing instrument can be characterized. 

Our own model is proof that a machine learning model can learn the relationship between the morphology parameters of a 2 ks and 10 ks observation. This suggests that machine learning models could characterize the relationship between morphology parameters derived from the observations of one instrument to those derived from the observations of another. With that in mind, we advise that this or any other follow-up merit assessment tool be developed with careful consideration for its intended use. One use case we envision for our tool is informing observers as to whether a \textit{eROSITA}-observed cluster has an extreme morphology relative to other \textit{eROSITA}-observed clusters (e.g., very asymmetric vs very symmetric). This information would be useful in determining whether a cluster has a dynamical state or core type that merits selection for follow-up. On the other hand, if one intends to predict \textit{Chandra}-resolution morphology parameters exactly, then one should train using \textit{Chandra}-resolution images as the truth images instead of \textit{eROSITA}-resolution images like we did.

\subsection{Alternative Metrics and Models}\label{Alternatives}

In our work, we aimed to create predictions that build on established techniques in the field of cluster cosmology. For computational and scientific reasons we opted for the simple but well established concentration, asymmetry, and smoothness parameters. While there is strong evidence of these parameters' correlation with our properties of interest \citep[e.g.,][]{Rasia_2013, Parekh_2015, Lovisari_2017, Green_2019}, these are not the only morphology parameters used in cluster cosmology and the statistical properties of these parameters are not well understood. This is especially true of the asymmetry parameter, where there exists plentiful literature on anisotropy tests and measures with better characterized statistical properties \citep[for examples and reviews of this topic see][]{mardia2000directional, feigelson2012modern, pewsey2013circular, baddeley2015spatial, RAJALA2018141}. While an analysis of the correlation of alternate anisotropy parameters to cluster properties of interest is outside the scope of this work, we encourage others to explore employing new measures of anisotropy to galaxy cluster cosmology.

Morphology is not the only potentially relevant observable for the outstanding questions in galaxy cluster cosmology. Other relevant properties such as X-ray scaling relations (e.g., Lx-Tx, Lx-M, Tx-M), X-ray surface brightness profiles, reconstruction of 3D gas and temperature profiles, or hydrostatic mass profiles might be of interest to X-ray observers. We chose morphology parameters because they are computationally easy to calculate (which is very relevant for training a model), applicable to observations of individual clusters, and are valid for the available simulation data. There is a wide parameter space of valid galaxy cluster properties to investigate, and moreover, our research is of potential interest to astronomy outside of X-ray galaxy cluster observation. We encourage others to explore this ample parameter space, but we opt to more fully explore what we view as the most promising avenue for success.

\begin{figure*}
    \centering
    \null \vspace{-20pt}
    \includegraphics[width=6in]{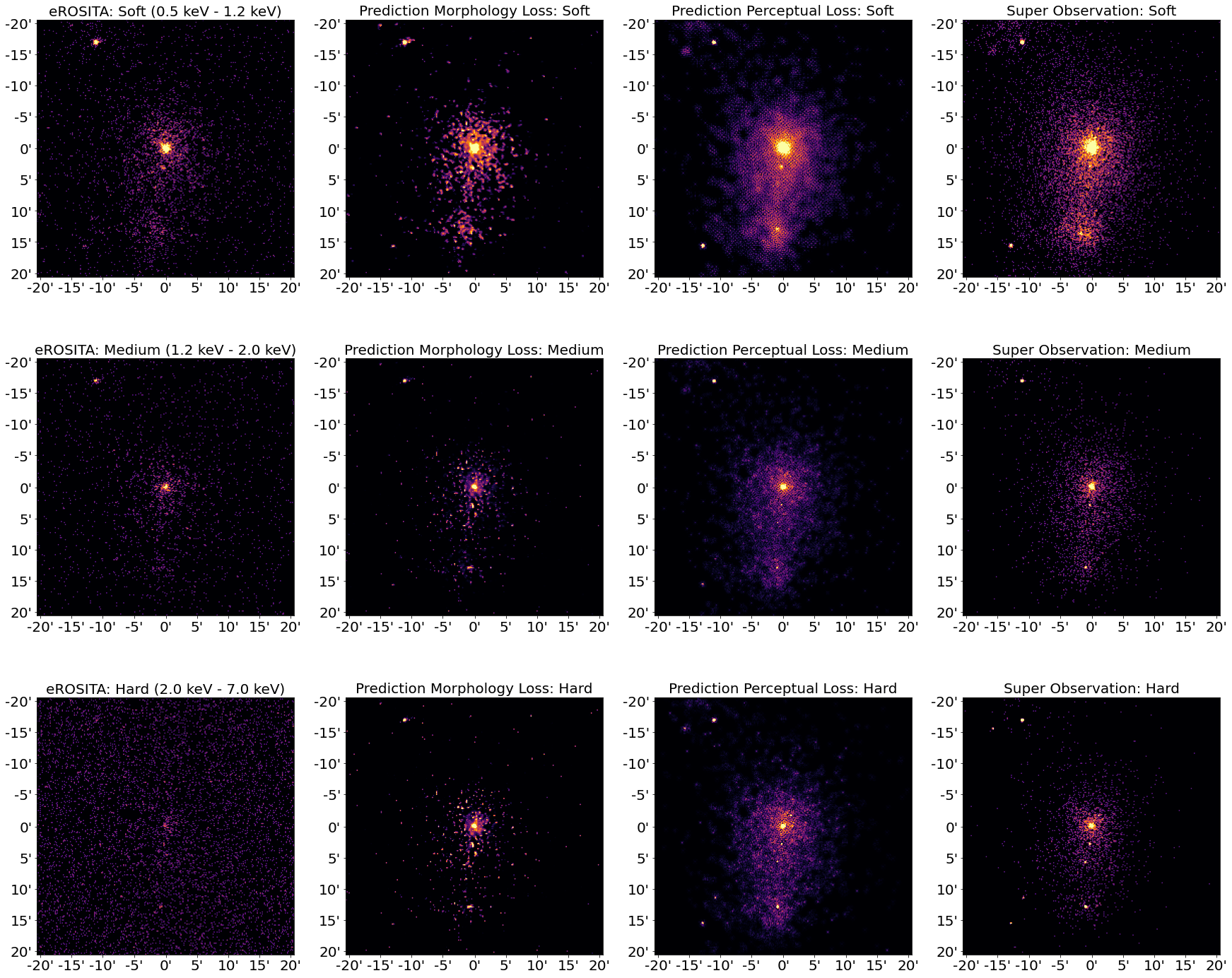}
    \caption{\raggedright Plots of different observations of the same cluster. From left to right: mock eROSITA inputs, morphology loss predictions, perceptual loss predictions, and the super observation targets. The perceptual loss images performed worse at matching the morphology parameters of the super observations, so we instead presented on our morphology loss model. The perceptual loss model produces more visually appealing, and arguably realistic looking, predictions than the morphology loss model, however. The perceptual loss model also offered superior accuracy in the soft band concentration parameter. The perceptual loss model introduces non-physical structures into the image and its predictions have less accurate asymmetry and smoothness relative to the morphology loss model. We note this to emphasize the important decisions one must make in designing, training, and evaluating an algorithm. Our model was designed with a specific use case in mind, but other follow-up merit assessment algorithms might choose different algorithm architecture or accuracy metrics.}
    \label{fig:loss_functions}
\end{figure*}

We chose to focus on the morphological accuracy of observations, and therefore designed our model appropriately for that purpose. Alternative algorithm architectures are also possible and ought to be considered depending on the priorities of the user. We considered different loss functions, including a perceptual loss function, inspired by \citet{ploss}, that uses the output of third layer of the VGG19 network \citep{VGG19}. An example prediction from that model is shown in Figure \ref{fig:loss_functions}. The perceptual loss function model produced prediction observations that better preserved concentration in the soft band than the morphology loss model. Moreover, the prediction images from the perceptual loss model are arguably more realistic, appearing smoother and less noisy, while clearly preserving substructure and AGN. However, the prediction observations did not preserve asymmetry or smoothness effectively, or preserve concentration in the hard band. We valued the preservation of asymmetry and smoothness by the morphology loss model higher than the improvement in concentration and appearance that the perceptual loss model offered, but we recognize that other observers might have different priorities. In addition, we considered using a standard UNET architecture \citep{UNET}, however it did not function well with our chosen loss function. The choice of evaluation metric, used to test the utility of the trained model, is also important. We chose to focus on morphology parameters, but many image-to-image machine learning models were designed instead to produce realistic-looking images that could fool human observers \citep[see][for examples of different image accuracy metrics, including human evaluation]{Dahl_2017}. Machine learning offers a wide parameter space in terms of algorithm hyperparameters and we do not argue that the model we have presented is the most optimal. Instead, we argue it illustrates that a machine learning approach to follow-up merit assessment is not only possible, but potentially more powerful than alternative solutions.

\section{Conclusion}\label{conclusion}
Galaxy clusters are an important probe of cosmology and useful laboratories for understanding physics. Key cluster properties, like core cooling and dynamical state, remain poorly understood and are in need of further study. Detailed follow up observations provide insight to these properties, but resources for follow-up are limited and an efficient method for evaluating the merit of these observations is needed. Machine learning offers one such method.

Our follow-up merit assessment tool can predict background-free, long duration, observations with accurate and precise morphology parameters. Given morphology’s correlation with the aforementioned cluster properties \textemdash{} which are both important in there own right and useful for selecting clusters for dark matter studies and mass estimation \textemdash{} morphologically accurate observations will aid follow-up selection. Our model will therefore advance our understanding of galaxy cluster internal physics, galaxy cluster cosmology, and cosmology more broadly.

Additionally, our work illustrates how follow-up merit assessment might be approached for a variety of different observational goals. Our model was designed to prioritize morphological accuracy, however the model could be trained as needed to address different properties. In working on this problem we explored a wide range of models, many with their own strengths and deficiencies. We advocate that observers strongly consider the priorities of their own observations when designing a follow-up merit assessment tool and tailor it to their specific needs. When utilized appropriately, machine learning can be an incredibly powerful tool for advancing galaxy cluster cosmology.

\acknowledgements
The material presented is based upon work supported by NASA under award No. 80NSSC22K0821. This research project was conducted using computational resources at the Maryland Advanced Research Computing Center (MARCC). We thank the MARCC support staff for invaluable help. We thank Christian Kirsch, Ole König, Maximilian Lorenz, and Joern Wilms for help installing and running the \texttt{SIXTE} software. We thank Veronica Biffi, Klaus Dolag, and Antonio Ragagnin for help running \texttt{PHOX}, using \texttt{Magneticum} data, and with the Cosmology Web Portal. We thank Tibor Rothschild for providing an example concentration fitting algorithm. We are grateful to Brianna Galgano for comments and suggestions important to this work.

\software{Astropy \citep{astropy:2013,astropy:2018}, Matplotlib \citep{matplotlib}, Python \citep{Python3}, Tensorflow \citep{tensorflow2015-whitepaper}, SIXTE \citep{SIXTE}.}

\newpage
\
\newpage

\bibliography{myref.bib}

\end{document}